\documentclass[num-refs]{ametsocV6.1}

\newcommand\BibTeX{{\rmfamily B\kern-.05em \textsc{i\kern-.025em b}\kern-.08em
		T\kern-.1667em\lower.7ex\hbox{E}\kern-.125emX}}

\usepackage{moreverb}

\usepackage[utf8]{inputenc}
\usepackage[T1]{fontenc}
\usepackage{amsmath}
\usepackage{comment}
\usepackage{esint}
\usepackage{amssymb}
\usepackage{graphicx}
\usepackage{placeins}
\usepackage{graphicx}
\usepackage{tikz,pgfplots}% Required for inserting images
\usetikzlibrary{positioning}
\usepackage[caption=false]{subfig}
\usepackage{float}
\usepackage{subeqnarray}
\usepackage{lipsum}
\usepackage{siunitx}
\usepackage{color}
\usepackage[T1]{fontenc}
\usepackage{epic}

\usepackage{eurosym}
\usepackage[colorinlistoftodos]{todonotes}
\usepackage{comment}

\newcommand{\be}{\begin{equation}}
	\newcommand{\ee}{\end{equation}}
\newcommand{\bae}{\begin{eqnarray}}
	\newcommand{\eae}{\end{eqnarray}}
\newcommand{\bse}{\begin{subeqnarray}}
	\newcommand{\ese}{\end{subeqnarray}}

\title{A Reduced-Dimensional Model for the Interhemispheric Geostrophic Meridional Overturning Circulation}
\clearpage
\authors{Elian Vanderborght\aff{a}\correspondingauthor{Elian Vanderborght, e.y.p.vanderborght@uu.nl}, and
	Henk A. Dijkstra \aff{a,b} }

\affiliation{\aff{a}Institute for Marine and Atmospheric research Utrecht, Department of Physics, 
	Utrecht University, Utrecht, the Netherlands\\
	\aff{b} Centre for Complex Systems Studies,  Utrecht University, Utrecht, the Netherlands. }
\abstract{The Global Overturning Circulation (GOC) is a key component of the climate system, transporting heat, carbon, and salt throughout the global ocean. Previous reduced-dimensional models have sought to represent this three-dimensional circulation but often neglected three key observational features: (1) the meridional overturning circulation is in geostrophic balance below the Ekman layer, (2) diapycnal mixing is strongly enhanced near ocean boundaries, and (3) upwelling is partly driven by adiabatic dynamics in the Southern Ocean. Building on \cite{callies2012simple}, we develop a reduced model that consistently incorporates all three by simulating temperature in latitude–depth space along the eastern and western boundaries of a semi-enclosed basin connected in the south to a zonally periodic re-entrant channel. The model clarifies how zonal temperature differences in the basin arise and are maintained through adiabatic and diffusive processes, giving rise to the geostrophic GOC. It also provides a transparent framework for understanding how geostrophic currents cross the equator to form the interhemispheric overturning, and how boundary-intensified mixing and Southern Ocean winds regulate polar downwelling rates. The reduced model shows good agreement with both a three-dimensional ocean model and theoretical scaling laws for stratification and overturning strength. Owing to its simplicity, it is well suited for long integrations exploring the GOC response under extreme forcing scenarios and offers a useful framework for testing eddy and mixing parameterizations.}
\begin{document}		
	\maketitle
	\statement{The Global Overturning Circulation (GOC) transports heat, salt, and carbon throughout the global ocean. To better understand its dynamics, previous studies have developed models that describe the circulation with a reduced number of degrees of freedom. However, these models often neglect that the GOC flow arises from a balance between the Coriolis force and a zonal pressure gradient, and that the mixing processes driving the circulation are strongly intensified near the boundaries. In this study, we develop a reduced model that accounts for these properties. The model provides insight into the three-dimensional structure of the circulation, its dependence on mixing strength and surface wind stress, and how such a circulation can cross the equator to become interhemispheric. Owing to its reduced nature, the model offers significant computational advantages for investigating the century- to millennium-scale response of the GOC to various forcing scenarios.}
	
	\newpage
	
	\section{Introduction}{\label{Intro}}
	The global overturning circulation (GOC) is the zonally integrated north–south circulation of the ocean. It combines wind-driven and buoyancy-driven components \citep{marshall2012closure, cessi2019global, roquet2025controls}, both ultimately powered by differential heating of Earth’s surface by solar radiation and tides. Differential heating drives large-scale atmospheric motions, where rising and sinking air masses generate wind stress at the ocean surface. This stress produces Ekman transport, which transfers surface buoyancy anomalies downward and forms the well-studied ventilated thermocline \citep{luyten1983ventilated, killworth1987continuously, pedlosky1996ocean}. In addition to this adiabatic thermocline, cross-diapycnal flows driven by small-scale turbulence establish a diffusive internal thermocline \citep{salmon1990thermocline, samelson1997large}. Together, these adiabatic and diabatic processes shape the stratification of the upper ocean.
	
	In an enclosed basin, zonal variations in upper-ocean stratification drive geostrophic meridional motions. Yet, for realistic diapycnal mixing rates, stratification is confined to the upper few hundred meters, and motions remain restricted to this shallow depth range. This picture conflicts with the significant role of the mid-depth and abyssal circulation in the GOC \citep{lumpkin2007global}. Since these large scale circulations are geostrophic in nature \citep{hirschi2003monitoring,johns2005estimating,waldman2021clarifying}, they too must be linked to zonal variations in stratification.
	
	For the mid-depth branch, this puzzle has been partly resolved by recognizing the influence of Southern Ocean winds \citep{toggweiler1995effect,toggweiler1998ocean,gnanadesikan1999simple}. In the absence of meridional continental boundaries, westerly winds drive an overturning circulation that penetrates to great depths. When mid-depth diffusivities are weak, dense North Atlantic Deep Water (NADW) is not mixed upward within the basin but instead flows southward and is adiabatically upwelled in the Southern Ocean \citep{lumpkin2007global,wolfe2011adiabatic,marshall2012closure}. As a result, the mid-depth basin stratification—and its zonal variation—is controlled by Southern Ocean dynamics \citep{wolfe2010sets,nikurashin2012theory}. %However, to our knowledge, the precise mechanism by which Southern Ocean dynamics communicate zonal variations to the basin stratification, thereby adiabatically closing the mid-depth circulation, remains unclear.
	
	The abyssal branch is supplied by Antarctic Bottom Water (AABW), which enters from the Southern Ocean, spreads northward along the seafloor, upwells diabatically within the basins, returns southward, and eventually upwells adiabatically to the Southern Ocean surface \citep{lumpkin2007global}. However, the low interior diapycnal mixing in the basin alone cannot explain its strength or density structure. Observations show that vertical diffusivity is strongly enhanced near boundaries, within a few hundred meters of the seafloor and continental slopes \citep{polzin1997spatial,st2002role,st2012turbulence}, driven by the breaking of internal waves \citep{wunsch2004vertical,garrett2007internal,nikurashin2013overturning}. This localized mixing drives strong upwelling along bottom slopes, thereby shaping the abyssal overturning circulation and its associated stratification \citep{ferrari2016turning,callies2018dynamics}.
	
	There is growing model-based evidence indicating that the GOC is expected to change under anthropogenic climate change \citep{levang2020causes,weijer2020cmip6,baker2025continued,vanwesten2025JGR}. Given the importance of geostrophic motions, these changes can be interpreted as arising from shifts in large-scale pressure differences. However, the complexity of comprehensive Global Climate Models (GCM) can obscure the individual processes that shape this response \citep{gerard2024diagnosing}. One way forward is therefore to consider reduced complexity models that capture the essential dynamics of the GOC.
	
	Simplified models of the GOC have been developed in earlier studies. However, they often fail to capture two key characteristics of the basin-scale circulation highlighted above: (1) the GOC is largely geostrophic, and (2) vertical diapycnal motions are concentrated near ocean boundaries. For example, ocean box models, which describe the ocean as a set of interconnected boxes in latitude–depth space, typically parameterize the overturning strength based on a meridional density contrast, effectively treating it as ageostrophic \citep{gnanadesikan1999simple,dijkstra2024role}. Furthermore, the response of such models to external forcing is highly sensitive to the specific form of this parameterization \citep{johnson2007reconciling,cimatoribus2014meridional}, introducing substantial ambiguity. Similarly, two-dimensional latitude–depth models of the GOC rely on non-geostrophic assumptions \citep{marotzke1988instability, wright1991zonally, sevellec2016amoc}. The zonal-mean nature of both model types implicitly assumes that diapycnal mixing—and the compensatory vertical motions it drives—is uniform in longitude, neglecting the strong boundary-localized mixing observed in the real ocean.
	
	\cite{marotzke1997boundary} showed that when vertical mixing is concentrated near ocean boundaries, vertical motions are likewise confined to regions adjacent to these boundaries. Although current understanding indicates that enhanced near-boundary mixing drives net upwelling along sloping, rather than strictly vertical, boundaries \citep{ferrari2016turning}, this framework still captures the key point: cross-isopycnal motions remain confined to a narrow region near the boundary where mixing is enhanced. Furthermore, restricting vertical mixing to the boundaries produces a zonally flat structure of isopycnals in the ocean interior \citep{welander1971discussion,mcdougall2017abyssal}, in agreement with observations \citep{hogg1999direct}. Building on \cite{marotzke1997boundary}, \cite{callies2012simple} developed a two-plane model in which interior isopycnals map to the eastern boundary density while all zonal temperature gradients are confined to a western boundary layer. The model reconstructs the three-dimensional ocean temperature field from the latitude–depth structure of temperatures at the eastern and western boundaries. Simulations of these boundary temperatures allow the two-plane model to reproduce the geostrophic overturning circulation, with results consistent with three-dimensional numerical ocean models \citep{marotzke1997boundary,scott2002location}.
	
	The model of \citet{callies2012simple} was formulated for a single hemisphere and included only diapycnal mixing as the upwelling source balancing northern downwelling. A natural extension of this framework is to formulate it for two hemispheres and include an adiabatic upwelling pathway. The first extension raises the question of how a geostrophic current can cross the equator. By geostrophic balance, a unidirectional cross-equatorial flow would require a reversal of the pressure gradient across the equator \citep{klinger1999behavior}—a striking feature that, to our knowledge, has not yet been represented in a reduced model of the GOC. The second extension involves introducing a zonally periodic, re-entrant southern channel forced by surface westerlies. This, in turn, prompts further questions: How do adiabatic channel dynamics enable interhemispheric flow with a reversing pressure gradient across the equator? And does the adjustment mechanism toward such an interhemispheric state fundamentally differ from the diffusive case?
	
	These questions motivate our study. Section~\ref{S:enclosed} begins by extending the \cite{callies2012simple} model to a two-hemisphere configuration. In this setting, we examine how the double-hemispheric two-plane model permits cross-equatorial transport and evaluate its consistency with the theory of \cite{marotzke2000dynamics}. In Section~\ref{S:Channel}, we extend the model to include a re-entrant channel, where wind-driven flows allow for deep adiabatic upwelling. Here, we place particular emphasis on the distinct nature of cross-equatorial flow that arises under adiabatic conditions. Throughout, we compare our results with those from a three-dimensional ocean model and theoretically derived scaling relations. Our results are discussed and summarized in Section~\ref{S:SumDis}
	
	\section{Interhemispheric flow in an enclosed basin}{\label{S:enclosed}}
	\subsection{Formulation and Model Domain}{\label{S:enclosed_Domain}}
	The Reduced-Geostropic-Global-Overturning-Circulation-Model (RGGOCM) is designed to simulate planetary-scale flows, which requires formulating all equations in spherical coordinates. As noted in Section~\ref{Intro}, these flows are predominantly geostrophic below the Ekman layer. Moreover, because the enclosed basin has meridional boundaries, wind-driven motions remain confined to the shallow, ventilated thermocline and are thus not expected to contribute significantly to mid-depth or abyssal stratification \citep{nikurashin2012theory}. Consequently, winds over the basin are neglected entirely.
	
	The geostrophic nature of these planetary flows implies a low Rossby number. Inserting this assumption into the Boussinesq equations on a rotating sphere yields the well-studied planetary geostrophic equations \citep{samelson2011theory}, consisting of two momentum equations, the hydrostatic relation, the continuity equation, and the thermodynamic equation, which respectively read:
	\begin{align}
		-fv &= -\frac{1}{a\cos(\theta)}\frac{\partial p}{\partial \lambda} - ru, \label{xmom_bas}\\
		fu &= -\frac{1}{a}\frac{\partial p}{\partial \theta} - rv, \label{ymom_bas}\\
		0 &= -\frac{\partial p}{\partial z} + g\alpha T, \label{zmom_bas}\\
		0 &= \frac{1}{a\cos(\theta)}\frac{\partial u}{\partial \lambda} + \frac{1}{a\cos(\theta)}\frac{\partial (v\cos\theta)}{\partial \theta} + \frac{\partial w}{\partial z}, \label{cont_bas}\\
		\frac{\partial T}{\partial t} + \frac{u}{a\cos(\theta)}\frac{\partial T}{\partial \lambda} + \frac{v}{a}\frac{\partial T}{\partial \theta} + w\frac{\partial T}{\partial z} &= \frac{\partial}{\partial z}\left(\kappa\frac{\partial T}{\partial z}\right) + \frac{1}{a^2\cos\theta}\frac{\partial}{\partial \theta}\left(\cos(\theta)\xi\frac{\partial T}{\partial \theta}\right)+c. \label{td_bas}
	\end{align}
	Here, $u$, $v$, and $w$ denote the Eulerian zonal, meridional, and vertical velocities, and $\lambda$, $\theta$, and $z$ are the zonal and meridional angular coordinates and the vertical coordinate. The Coriolis parameter is $f = 2\Omega \sin\theta$, with $\Omega = 7.2\times10^{-5}$ s$^{-1}$ the Earth's rotation rate; $r$ is a linear Rayleigh friction coefficient; and $a = 6400$ km is the Earth's radius. 
	
	\citet{callies2012simple} proposed using equatorial thermal wind relations, which are more consistent with observations \citep{lukas1984geostrophic, lagerloef1999tropical}. However, these relations tend to generate numerical instabilities near the equator, likely because nonlinear and frictional effects are required to maintain stable flow. Introducing linear Rayleigh friction stabilizes the solution and is therefore adopted here, while its influence is kept minimal by ensuring $r \ll f$ at latitudes sufficiently far from the equator.
	
	The hydrostatic relation (\ref{zmom_bas}) links dynamic pressure $p$ to the fluid density, which here depends solely on temperature $T$, and relates to buoyancy via the linear equation of state $b = g\alpha T$, where $g=9.81$ m s$^{-2}$ is gravitational acceleration and $\alpha=2\times10^{-4}$ °C$^{-1}$ is the thermal expansion coefficient.
	
	Diapycnal mixing is represented by a vertical diffusivity $\kappa$. Although the cross-isopycnal direction is not always perfectly vertical, such an  approximation is commonly adopted. Horizontal mixing is primarily included for numerical stability and, for the values of the diffusivity $\xi$ considered here, does not significantly affect the solution (not shown). 
	
	Finally, the time evolution of temperature may produce static instabilities. These are removed using the Rahmstorf convective adjustment scheme \citep{rahmstorf1993fast}, which introduces the tendency term $c$ in equation~(\ref{td_bas}).
	
\begin{figure}
		\captionsetup{justification=centering}
		\centering
		{\includegraphics[width=0.5\textwidth]{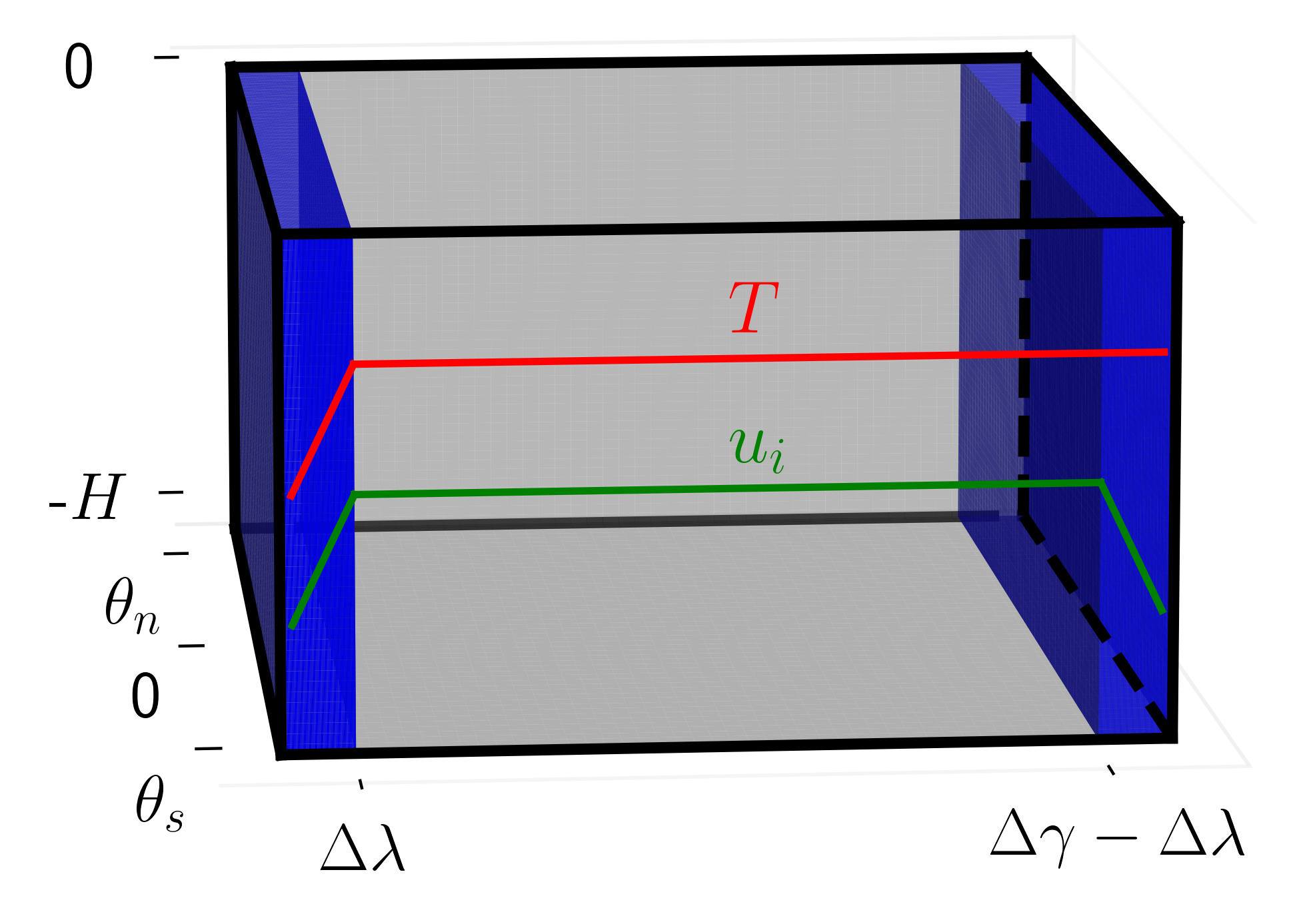}}
		\caption{\em \fontsize{10.5}{12.5}\selectfont Schematic of the model domain. Blue regions correspond to western and eastern boundary layer and have a zonal width of $\Delta \lambda$. The red line represents typical structure of zonal temperature profile, with zonal gradients confined to the western boundary layer. Green line represents zonal velocity, which is constant over the interior and decays to zero within the boundary layer.}
		\label{F:Fig1}
	\end{figure}

	The model domain for solving equations~(\ref{xmom_bas})–(\ref{td_bas}) is illustrated in Fig.~\ref{F:Fig1}. It spans from $\theta_s = -70^\circ$S to $\theta_n = 70^\circ$N in latitude, $0^\circ$E to $\Delta\gamma = 60^\circ$E in longitude, and from the surface ($z = 0$) to the ocean bottom ($z = -H = -4$ km).
	
	\subsection{Assumptions and Governing Equations}{\label{S:Assumptions_enclosed_Domain}}
	The assumptions underlying the RGGOCM were previously outlined by \cite{callies2012simple}. Here, we summarize the key elements and present the governing equations, while referring the reader to \cite{callies2012simple} for a more detailed discussion.
	
	The model domain (Fig.~\ref{F:Fig1}) is divided into three regions: an eastern boundary layer, a western boundary layer, and the ocean interior. The boundary layers have a characteristic zonal thickness of $\Delta\lambda$, within which mixing is assumed to occur, while the interior is characterized by zero diapycnal mixing. In the absence of diapycnal mixing, equations (\ref{xmom_bas})–(\ref{zmom_bas}) together with (\ref{td_bas}) reduce to the so-called thermocline, or $M$-equation \citep{welander1959advective, welander1971discussion}, which admits the trivial solution of a zonally uniform temperature field.
	%\footnote{\citet{welander1959advective} derived this result for a fully geostrophic, adiabatic flow without Rayleigh friction. \citet{samelson1997simple} extended the derivation to include linear Rayleigh friction, showing that even in the adiabatic case a zonally uniform temperature field is no longer strictly a solution. However, under the assumption $r \ll f$, the temperature field is expected to remain nearly zonally flat to leading order in the adiabatic limit.} 
	Such zonal flatness of interior isopycnals is also consistent with observations \citep{hogg1999direct}.
	
	Developments in our understanding of thermohaline circulation adjustment suggest that buoyancy anomalies are transmitted across the basin by Kelvin and Rossby waves. The initial, nearly instantaneous response is mediated by boundary and equatorial Kelvin waves, which remove pressure gradients along the equator and the eastern boundary. This is followed by the slower westward propagation of long Rossby waves from the eastern boundary, which adjust the thermocline depth in the interior toward that of the east \citep{johnson2002theory, marshall2013propagation}. On long time scales, this process produces a zonally uniform interior temperature field, set by the eastern boundary. Near the western boundary, however, Rossby waves cannot eliminate zonal gradients. These gradients are instead confined within a narrow western boundary layer (Fig.~\ref{F:Fig1}), where the temperature varies approximately linearly from the eastern boundary value ($T_e$) in the interior to the western boundary value ($T_w$). Thus, the basin-wide temperature field is fully determined by $T_w(z,\theta)$ and $T_e(z,\theta)$, and the prognostic equations for these two variables govern the system’s evolution.
	
	In the interior ocean, a zonal flow may arise from meridional gradients in the eastern boundary temperature. Assuming zonally flat isopycnals in the interior, combining equations~(\ref{xmom_bas})–(\ref{zmom_bas}) yields an expression for the vertical shear of the interior zonal flow:
	\begin{equation}
		\frac{\partial u_i}{\partial z}=-\frac{\alpha g f}{a(f^2+r^2)}\frac{\partial T_e}{\partial \theta}.
		\label{ui}
	\end{equation}
	Near the eastern boundary, however, the flow must turn ageostrophic to satisfy the no-normal-flow condition. \cite{cessi2009eddy} suggested that eddy-driven zonal circulation can precisely cancel the Eulerian zonal geostrophic flow close to the boundary. As a result, the residual zonal flow (Eulerian plus eddy) vanishes, even though each component may be nonzero. This mechanism allows meridional temperature gradients along the boundary without violating the no-normal-flow condition. To parameterize this opposing effect of the eddy-driven  circulation, we let $u_i$ linearly decrease towards zero in both the eastern and western boundary layer (Fig.~\ref{F:Fig1}).
	
	Zonal temperature gradients are confined near the western boundary, giving rise to a geostrophic meridional flow within the western boundary layer. By combining equations~(\ref{xmom_bas})–(\ref{zmom_bas}) to obtain an expression for the vertical shear of $v$ and averaging over the western boundary layer, we find $v_w$ from:
	\begin{equation}
		\frac{\partial v_w}{\partial z}=\frac{\alpha g}{f^2+r^2}\left(\frac{f}{a\cos(\theta)\Delta\lambda}(T_e-T_w)-\frac{r}{2a}\frac{\partial}{\partial \theta}(T_e+T_w)\right),\label{vw}
	\end{equation}
	A similar expression can be derived for the eastern boundary meridional velocity, which is purely frictional. Since $r \ll f$, this flow is negligible compared to $v_w$ except very close to the equator. Moreover, one can  show that $\partial_\theta T_e \approx 0$ at the equator, making the ageostrophic meridional flow along the eastern boundary negligible even there. Its contribution is therefore neglected altogether.
	
	Averaging equation~(\ref{cont_bas}) over the eastern and western boundary layers, and noting that the meridional velocity along the eastern boundary vanishes, we obtain the following continuity equations for the eastern and western boundaries, respectively:
	\begin{align}
		\frac{\partial w_e}{\partial z}-\frac{u_i}{a\cos(\theta)\Delta\lambda}&=0,\label{conte}\\
		\frac{\partial w_w}{\partial z}+\frac{1}{a\cos(\theta)}\frac{\partial (v_w\cos(\theta))}{\partial\theta}+\frac{u_i}{a\cos(\theta)\Delta\lambda}&=0,\label{contw}
	\end{align}
	Note that the structure of the flow field implies that vertical velocities are confined within the narrow boundary layers. This agrees with GCM boundary mixing simulations from \cite{marotzke1997boundary} and \cite{scott2002location}. 
	
	We may combine equations (\ref{conte})-(\ref{contw}) to define a basin overturning streamfunction $\psi_b$ as:
	\begin{align}
		-\frac{\partial \psi_b}{\partial z}=a \cos(\theta)\Delta\lambda v_w,  && 	\frac{\partial \psi_b}{\partial \theta}=a^2\cos(\theta)\Delta\lambda (w_e+w_w).
		\label{psib}
	\end{align}
	Applying all assumptions to the thermodynamic equations~(\ref{td_bas}), the temperature equations at the eastern and western boundary read:
	\begin{align}
		\frac{\partial T_e}{\partial t}+w_e\frac{\partial T_e}{\partial z}&=\frac{\partial}{\partial z}\left(\kappa_b\frac{\partial T_e}{\partial z}\right)+\frac{1}{a^2\cos(\theta)}\frac{\partial}{\partial \theta}\left(\xi_b\cos(\theta)\frac{\partial T_e}{\partial \theta}\right)+c_e,\label{Te}\\
		\frac{\partial T_w}{\partial t}+\frac{v_w}{a}\frac{\partial T_w}{\partial \theta}+w_w\frac{\partial T_w}{\partial z}&=\frac{\partial}{\partial z}\left(\kappa_b\frac{\partial T_w}{\partial z}\right)+\frac{1}{a^2\cos(\theta)}\frac{\partial}{\partial \theta}\left(\xi_b\cos(\theta)\frac{\partial T_w}{\partial \theta}\right)+c_w\label{Tw}.
	\end{align}
	Note that $\kappa_b$ and $\xi_b$ represent the value of the mixing coefficients within the boundary layer.  Elsewhere, these coefficients are assumed to be zero. 
	
	While \cite{callies2012simple} included the effect of Rossby wave radiation, they did not account for the initial Kelvin wave adjustment. Boundary Kelvin waves tend to meridionally flatten the eastern boundary temperature, but we will show below that meridional flatness is automatically satisfied in steady state. Therefore, an explicit treatment of boundary Kelvin waves may be redundant. In contrast, eastward-propagating equatorial Kelvin waves remove zonal pressure gradients along the equator \citep{johnson2019recent}. This effect is not captured in equations~(\ref{Te})–(\ref{Tw}), and we include it by a Kelvin-wave adjustment parameterization by  instantaneously relaxing $T_e(z,0^\circ)$ to $T_w(z,0^\circ)$.

	\subsection{Boundary conditions}
\begin{figure}
		\captionsetup{justification=centering}
		\centering
		{\includegraphics[width=0.5\textwidth]{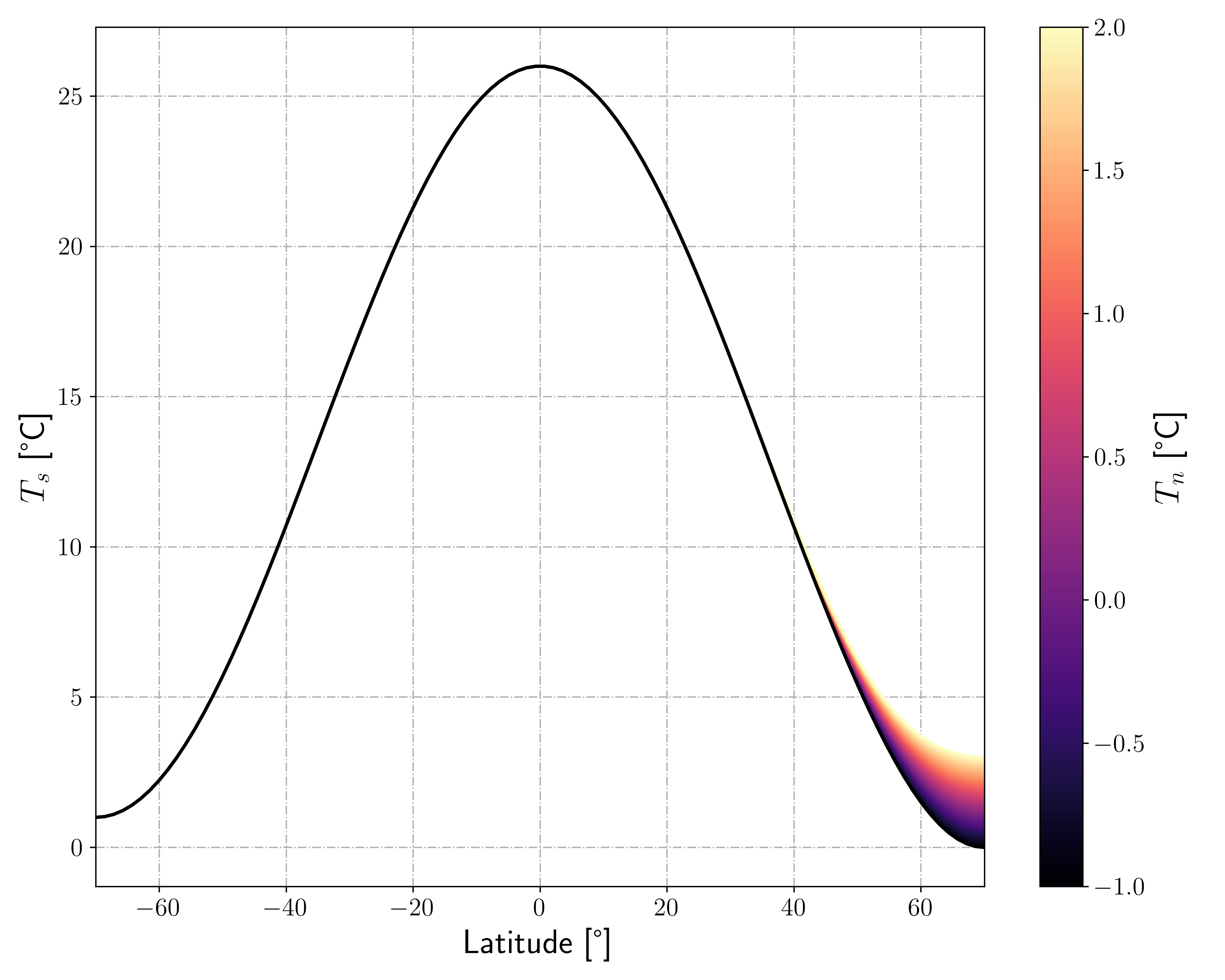}}
		\caption{\em \fontsize{10.5}{12.5}\selectfont Temperature relaxation profile (equation~(\ref{Ts})) for $\Delta T=25$°C, $T_{\mathrm{min}}=1$°C, $\delta_T=$ 1800 km and different values of $T_n$ (colors). }
		\label{F:Fig2}
	\end{figure}

{\label{S:BC_enclosed_Domain}}
	At the ocean bottom, $z=-H$, we impose a no-vertical-flux boundary condition, such that $\partial_z T_e(-H,\theta) = \partial_z T_w(-H,\theta) = 0$ for all $\theta$. At the surface, a flux boundary condition relaxes the upper-layer temperature toward a reference profile $T_s$, expressed as:
	\begin{align}
		\kappa_b\frac{\partial T_e}{\partial z} = \frac{D}{\mu}(T_s - T_e), &&
		\kappa_b\frac{\partial T_w}{\partial z} = \frac{D}{\mu}(T_s - T_w),
		\label{TopBC_T}
	\end{align}
	where $D$ is the mixed layer depth and $\mu$ a relaxation timescale. We choose $\mu$ sufficiently small so that the surface temperature closely follows the prescribed profile $T_s$. The relaxation profile is defined as:
	\begin{equation}
		T_s(\theta) = \frac{\Delta T}{2} \Bigg[\cos\left(\pi\frac{\theta}{\theta_n}\right) + 1\Bigg] + T_n e^{-(\theta-\theta_n)^2/\eta^2} + T_{\mathrm{min}}.
		\label{Ts}
	\end{equation}
	Here, $\Delta T$ sets the equator-to-pole temperature difference, and the second term introduces a hemispheric asymmetry. Following \cite{wolfe2014salt}, we set $\eta = 18$°. The parameter $T_{\mathrm{min}}$ defines the minimal temperature in the Southern Hemisphere (SH). Fig. \ref{F:Fig2} illustrates the relaxation profile for $\Delta T = 25$°C, $T_{\mathrm{min}}=1$°C and various values of $T_n$.
	
	A no-normal-flow condition at the ocean bottom and the rigid-lid approximation at the surface are imposed, requiring; $w_e(0,\theta) = w_e(-H,\theta) = w_w(0,\theta) = w_w(-H,\theta) = 0$. This condition is met when the vertically integrated zonal $u_i$ and meridional transport equals zero (equations~(\ref{conte})-(\ref{contw})). A no-normal-flow condition is also applied at the northern and southern boundaries of the domain. This can be imposed by requiring:
	\begin{align}
		T_e=T_w, && \frac{\partial T_e}{\partial \theta}+\frac{\partial T_w}{\partial \theta}=0,  && \text{ for $\forall z$ and $\theta=\theta_s, \theta_n$}
		\label{basin_no_normal}
	\end{align}
	Condition~(\ref{basin_no_normal}) implies that $w_e \approx w_w$ for all $z$ and at $\theta = \theta_s, \theta_n$, ensuring that the tendencies of $T_e$ and $T_w$ approximately match at these latitudes. This follows from substituting the definitions of $v_w$ (equation~(\ref{vw})) and $u_i$ (equation~(\ref{ui})) into the continuity equations~(\ref{conte})–(\ref{contw}) and applying the boundary conditions~(\ref{basin_no_normal}). The resulting expression shows that the difference between $w_e$ and $w_w$ scales with the parameter $r$, which, when chosen sufficiently small, makes this difference negligible.
	
	Equations (\ref{ui})-(\ref{Tw}) together with the boundary conditions, form a closed system. As analytic solutions, if at all possible,  cannot be easily derived, we rely on a numerical implementation, which is outlined in Appendix A. 
	
	\subsection{Asymmetric reference case}{\label{S:Sol_enclosed_Domain}}
	To study the interhemispheric overturning circulation in our model, we present results from a reference case  under an asymmetric forcing scenario with $T_n=-1$°C. The standard values of model parameters for the reference case are given in Table~\ref{T:Tab1}. Following \cite{marotzke2000dynamics}, the simulation is initialized from a steady-state solution obtained under symmetric forcing ($T_n=0$°C). 
	
	\begin{table}[ht]
		\centering
		\begin{tabular}{l c c}
			\hline
			Parameter & Symbol & Value \\ 
			\hline
			Boundary layer width & $\Delta \lambda$ & 4°\\
			Rayleigh friction parameter & $r$ & $4\times 10^{-6}$ s$^{-1}$\\
			Vertical diffusivity at boundary & $\kappa_b$ & $3 \times 10^{-4} \ \mathrm{m^2} \ \mathrm{s^{-1}}$ \\
			Horizontal diffusivity & $\xi_b$ & $2\times 10^3 \ \mathrm{m^2} \ \mathrm{s^{-1}}$ \\
			Mixed layer depth & $D$ & $50 \ \mathrm{m}$ \\
			Relaxation timescale & $\mu$ & $15 \ \mathrm{days}$ \\
			Equator-Pole temperature difference & $\Delta T$ & $25$°C\\
			Minimal temperature of SH & $T_{\mathrm{min}}$ & $1$°C\\ 		
			\hline
		\end{tabular}
		\caption{Model parameters used in reference case with asymmetric forcing.}
		\label{T:Tab1}
	\end{table}
	
\begin{figure}
		\captionsetup{justification=centering}
		\centering
		{\includegraphics[width=\textwidth]{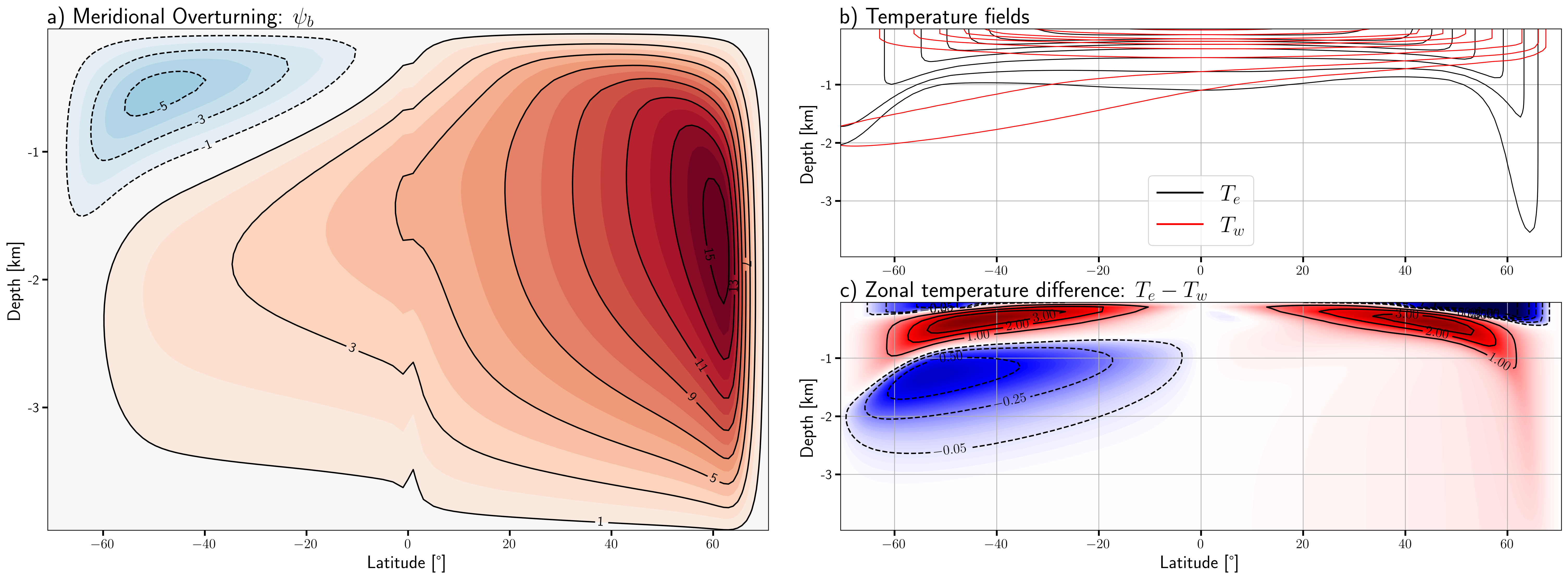}}
		\caption{\em \fontsize{10.5}{12.5}\selectfont Steady-state solution of the reference case: (a) Overturning streamfunction $\psi_b$ (equation~(\ref{psib})) in Sv. (b) Eastern and western boundary temperatures, with contours at [0.6, 1, 2, 4, 6, 8, 10, 15, 23] $^\circ$C. (c) Zonal temperature difference in $^\circ$C. In all panels, red shading indicates positive values, and blue shading indicates negative values.}
		\label{F:DH_MOC}
	\end{figure}

	Fig.~\ref{F:DH_MOC} shows the steady-state fields of the reference case under asymmetric forcing. In line with \cite{marotzke2000dynamics}, slightly stronger polar cooling in the Northern Hemisphere (NH) drives a positive asymmetric Northern sinking Overturning Circulation (NOC), with $5.2$~Sv crossing the equator and a maximum northward transport of $15.7$~Sv. A negative Southern sinking Overturning Circulation (SOC) persists in the Southern  Hemisphere (SH) but is substantially weaker, reaching a maximum southward transport of $5.6$~Sv. This corresponds to about $10.5$~Sv of upwelling in each hemisphere. The total upwelling of $21$~Sv is therefore nearly identical to that obtained under symmetric forcing (not shown).
	
	For discussion, we distinguish between thermocline isotherms, which outcrop in both hemispheres, and subthermocline isotherms, which outcrop only in the NH. Fig.~\ref{F:DH_MOC}b shows that both western and eastern boundary thermocline isotherms are approximately symmetric about the equator. However, the western boundary thermocline is more strongly stratified than the east, resulting in a positive zonal temperature difference across the thermocline (Fig.~\ref{F:DH_MOC}c). In contrast, while eastern boundary subthermocline isotherms remain nearly symmetric except at high latitudes, western boundary subthermocline isotherms slope gradually upward toward their NH outcrop latitude, intersecting the eastern boundary subthermocline isotherms at the equator. This produces a negative subthermocline temperature difference in the SH that reverses sign across the equator. 
	
\begin{figure}
		\captionsetup{justification=centering}
		\centering
		{\includegraphics[width=\textwidth]{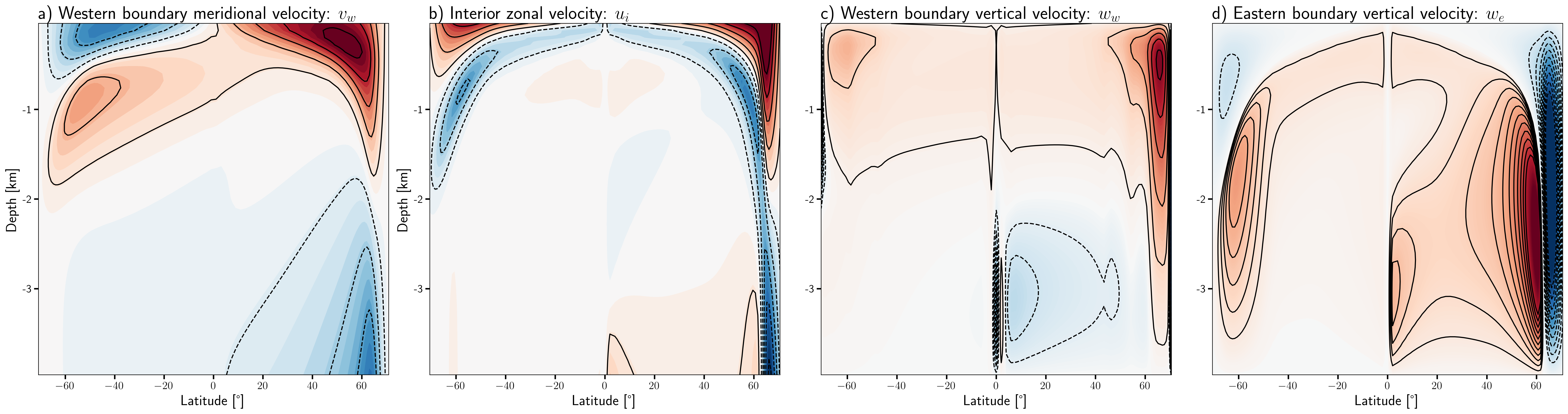}}
		\caption{\em \fontsize{10.5}{12.5}\selectfont Steady-state of the reference case  performed in the MITgcm: (a) Western boundary meridional velocity, with contour intervals $1.8$ cm s$^{-1}$. (b) Interior zonal velocity, with contour intervals $0.4$ cm s$^{-1}$. (c) Western boundary vertical velocity with contour intervals $6\times 10^{-5}$ cm s$^{-1}$ for negative values and $3\times 10^{-4}$ cm s$^{-1}$ for positive values. (d) Eastern boundary vertical velocity with contour intervals $1\times 10^{-3}$ cm s$^{-1}$ for negative contours and $1\times 10^{-4}$ cm s$^{-1}$ for positive values. In all panels, red shading indicates positive values, and blue shading indicates negative values.}
		\label{F:DH_vel}
	\end{figure}

	Along the eastern boundary thermocline, unstratified waters overlie stratified waters (Fig.~\ref{F:DH_MOC}b). From equation~(\ref{Te}), a transition from increasing to decreasing stratification toward the surface requires downwelling above upwelling, consistent with Fig.~\ref{F:DH_vel}d. This vertical structure corresponds to a three-layer zonal flow: eastward near the surface and bottom, separated by a westward intermediate layer (Fig.~\ref{F:DH_vel}b). Along the western boundary thermocline, upwelling dominates, but its magnitude is substantially smaller than the eastern boundary downwelling (Fig.~\ref{F:DH_vel}c). The opposing vertical motions along the boundaries sustain the contrasting thermocline stratification (Fig.~\ref{F:DH_MOC}c). The resulting thermocline temperature difference drives poleward motion in the surface layers (Fig.~\ref{F:DH_vel}a), sustaining the upper branch of the NH NOC and SH SOC.
	
	As the western boundary upwelling gradually intensifies poleward, the poleward surface flow strengthens (Fig.~\ref{F:DH_vel}a,c). At the polar boundary, the current turns eastward, sinks along the eastern boundary, and returns equatorward as a Deep Western Boundary Current (DWBC) (Fig.~\ref{F:DH_vel}b,d). The strong polar downwelling along the eastern boundary weakens local stratification (Fig.~\ref{F:DH_MOC}b) and warms the water column to great depth, while convective mixing along the western boundary removes heat from the poleward current. As a result, the high-latitude zonal temperature difference remains positive approximately down to the depth of eastern boundary downwelling, with a shallow negative anomaly at the surface caused by northward heat transport (Fig.~\ref{F:DH_vel}c).
	
	Clear asymmetries across the equator are evident in Fig.~\ref{F:DH_MOC} and~\ref{F:DH_vel}. The DWBC originating in the NH advects cold western boundary anomalies southward, maintaining a positive NH subthermocline temperature difference (Fig.~\ref{F:DH_MOC}c). After descending slightly along the western boundary (Fig.~\ref{F:DH_vel}c), the DWBC crosses the equator and supplies the SH eastern boundary upwelling through an eastward bottom current (Fig.~\ref{F:DH_vel}a,c,d). This upwelling lifts subthermocline isotherms along the SH eastern boundary, enhancing local stratification and shoaling the high-latitude mixed layer (Fig.~\ref{F:DH_MOC}b). The resulting reduction in the meridional temperature gradient weakens the eastward surface flow, leading to shallower and weaker eastern boundary downwelling and western boundary upwelling in the high-latitude SH (Fig.~\ref{F:DH_vel}b–d). Consequently, stratification along the SH western boundary weakens below the thermocline (Fig.~\ref{F:DH_MOC}b). The east–west asymmetry in SH boundary stratification therefore maintains a negative subthermocline temperature difference (Fig.~\ref{F:DH_MOC}c), which geostrophically sustains the cross-equatorial flow of the NOC.
	
\begin{figure}
		\captionsetup{justification=centering}
		\centering
		{\includegraphics[width=\textwidth]{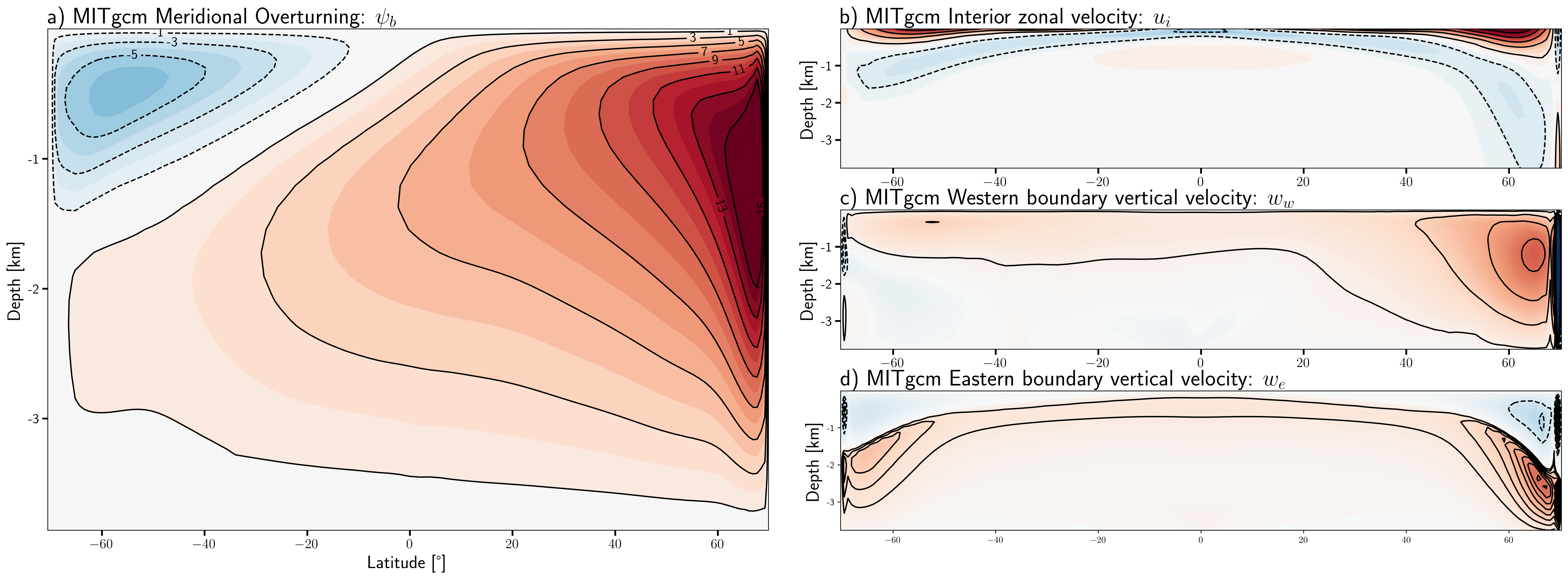}}
		\caption{\em \fontsize{10.5}{12.5}\selectfont Steady state of the reference experiment under asymmetric forcing, simulated using MITgcm: (a) overturning streamfunction computed from the zonally integrated meridional transport; (b) interior zonal velocity at 30°E; (c) western-boundary and (d) eastern-boundary vertical velocity, obtained as the longitudinal mean over a 4° band adjacent to each boundary. Contour intervals are identical to those in Figs. \ref{F:DH_MOC} and \ref{F:DH_vel}}
		\label{F:DH_MIT}
	\end{figure}

	Fig.~\ref{F:DH_MIT} shows results from a similar numerical experiment performed with the Massachusetts Institute of Technology general circulation model (MITgcm) \citep{marshall1997finite,marshall1997hydrostatic}. A more detailed presentation of the numerical setup is provided in Appendix B. The MITgcm results share many similarities with those of the RGGOCM. In particular, the overturning streamfunction responds asymmetrically to a weakly asymmetric heat flux. The interhemispheric NOC reaches a maximum northward transport of $20$~Sv, of which $5.2$~Sv crosses the equator, while the weaker SOC has a maximum southward transport of $6.9$~Sv. As in the RGGOCM, the eastern and western boundary thermocline are marked by upwelling and downwelling, respectively (Fig.~\ref{F:DH_MIT}c–d). This structure produces a positive zonal temperature difference across the thermocline (not shown, but similar to Fig.~\ref{F:DH_MOC}c) and drives the northward (southward) surface transport in the NH (SH).
	
	The northward surface transport converges in the high-latitude NH, feeding an eastward surface current (Fig. \ref{F:DH_MIT}b) that sinks along the eastern boundary and returns to the abyssal western boundary through a westward current. The DWBC then crosses the equator and supplies eastern boundary upwelling in the SH. As in the RGGOCM, this circulation upwelling leads to a substantially weaker and shallower high-latitude SH eastward surface current, along with reduced eastern boundary downwelling and western boundary upwelling (Fig. \ref{F:DH_MIT}b–c). Consequently, the global volume budget closely resembles that obtained with the RGGOCM.
	
	A difference between the two models lies in the location of western boundary downwelling. As noted by \cite{Marotzke2000}, the DWBC supplies western boundary downwelling after crossing the equator. Although very small, this is evident in the MITgcm (Fig.~\ref{F:DH_MIT}c) but occurs farther north in the RGGOCM, before the DWBC crosses the equator (Fig.~\ref{F:DH_vel}b). The discrepancy reflects differences in interior zonal flow: in the MITgcm, an eastward bottom current develops between $50$°S and the equator, whereas no such current forms in the RGGOCM. However, SH western boundary downwelling is not a characteristic feature of cross-equatorial flow in the MITgcm. When boundary–interior buoyancy exchange is reduced, for example by lowering viscosity or mesoscale diffusivity, the downwelling disappears (not shown). Its absence in the RGGOCM can therefore be attributed to the decoupling of boundary and interior regions.
	
	\subsection{Adjustment towards an asymmetric overturning state}\label{S:DH_Adjust}
	In Section~\ref{S:enclosed}\ref{S:Assumptions_enclosed_Domain}, we introduced a parameterized representation of equatorial pressure gradient adjustment through Kelvin wave propagation. To test its role, we repeated the experiment from Section~\ref{S:enclosed}\ref{S:Sol_enclosed_Domain} but excluded this adjustment. Without it, the RGGOCM is unable to simulate cross-equatorial flow and, moreover, becomes numerically unstable even under very weak asymmetric forcing (not shown). The sensitivity of the overturning circulation to this seemingly minor parameterization is striking. To investigate why equatorial adjustment is essential for enabling cross-equatorial flow, we follow the approach of \cite{Marotzke2000} and analyze the first 60 years of spin-up under asymmetric forcing, starting from a symmetrically forced steady state.
	
\begin{figure}
		\captionsetup{justification=centering}
		\centering
		{\includegraphics[width=\textwidth]{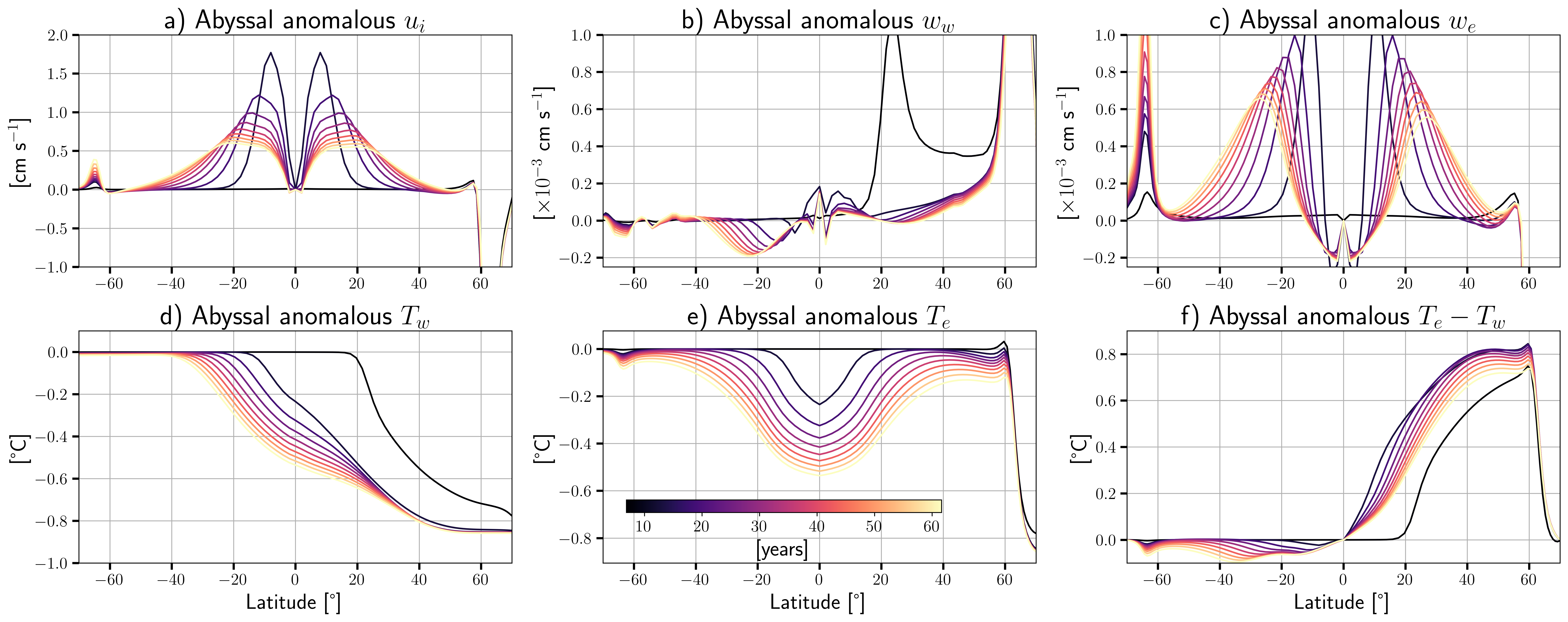}}
		\caption{\em \fontsize{10.5}{12.5}\selectfont Spin-up toward the reference case. All anomalies are calculated relative to the symmetrically forced steady state. (a) Anomalous zonal velocity averaged over the deepest $500$~m. Anomalous vertical velocity at the (b) eastern and (c) western boundaries, averaged over the full depth. Anomalous temperature at the (d) western and (e) eastern boundaries, and (f) their difference (eastern minus western), averaged over the deepest $2000$~m. Different line colors indicate different model times as shown in the color bar in panel (e).}
		\label{F:SpinUp}
	\end{figure}

	Imposing $T_n = -1$°C cools both the eastern and western boundaries of the NH. These anomalies increase the meridional temperature gradient, which is rapidly transmitted to the abyss through convective mixing (Fig.~\ref{F:SpinUp}d–e). The enhanced gradient intensifies the surface eastward flow and abyssal westward flow (Fig.~\ref{F:SpinUp}a). This, in turn, amplifies abyssal western boundary upwelling (Fig.~\ref{F:SpinUp}b) and eastern boundary downwelling (Fig.~\ref{F:SpinUp}c) at high northern latitudes. The cold signal and anomalous upwelling are carried southward along the western boundary by the DWBC, reinforcing the NH temperature contrast $T_e-T_w$ and reaching the equator after roughly 15 years (Fig.~\ref{F:SpinUp}f).
	
	Upon reaching the equator, the western boundary cold anomaly is transmitted to the eastern boundary via Kelvin wave–mediated temperature relaxation. This produces a sharp, bell-shaped cold anomaly at the eastern boundary (Fig.~\ref{F:SpinUp}e). The anomaly drives eastward flow on both sides of the equator (Fig.~\ref{F:SpinUp}a), which induces eastern boundary upwelling (Fig.~\ref{F:SpinUp}c). At the same time, it reduces anomalous western boundary upwelling in the NH and generates anomalous western boundary downwelling in the SH (Fig.~\ref{F:SpinUp}b). This results in warming of the abyssal west relative to the east. The eastward flow subsequently spreads northward and southward (Fig.~\ref{F:SpinUp}a), producing a symmetric expansion of the bell-shaped eastern boundary cold anomaly (Fig.~\ref{F:SpinUp}e) and a slowdown of the southward propagating cold front along the western boundary (Fig.~\ref{F:SpinUp}d). This allows the eastern boundary anomaly to catch up with the western boundary anomaly, eventually reducing $T_e-T_w$ anomaly across both hemispheres (Fig.~\ref{F:SpinUp}f).
	
	The decrease in $T_e - T_w$ produces a negative subthermocline zonal temperature anomaly in the SH (Fig.~\ref{F:SpinUp}f). This anomaly enables the DWBC to cross the equator, carrying cold northern-sourced water into the SH abyss. Upon reaching high southern latitudes, the DWBC turns eastward and feeds eastern boundary upwelling (Fig.~\ref{F:DH_vel}a,c). The associated upwelling strengthens stratification along the eastern boundary and, through the resulting shallower eastern boundary mixed layer, reduces upwelling and consequently weakens subthermocline stratification along the western boundary. In this way, the negative zonal temperature gradient (Fig.~\ref{F:DH_MOC}c) is maintained by the cross-equatorial flow of the DWBC.
	
	From this discussion, it is evident that up- and downwelling anomalies, which are necessary for the equatorial DWBC to cross the equator, can only occur if western boundary temperature anomalies are communicated to the eastern boundary. While \cite{kawase1987establishment} and \cite{marotzke2000dynamics} reported similar findings, the former emphasized the role of Kelvin waves in setting the adjustment timescale, whereas the latter found timescales more consistent with advective transport along the boundaries. The RGGOCM represents this adjustment as a combination of the two mechanisms, with Kelvin waves governing the equatorial crossing timescale and advection controlling the boundary transport.
	\subsection{Scaling of the asymmetric overturning circulation}\label{S:Scaling_DH}
\begin{figure}
		\centering
		{\includegraphics[width=\textwidth]{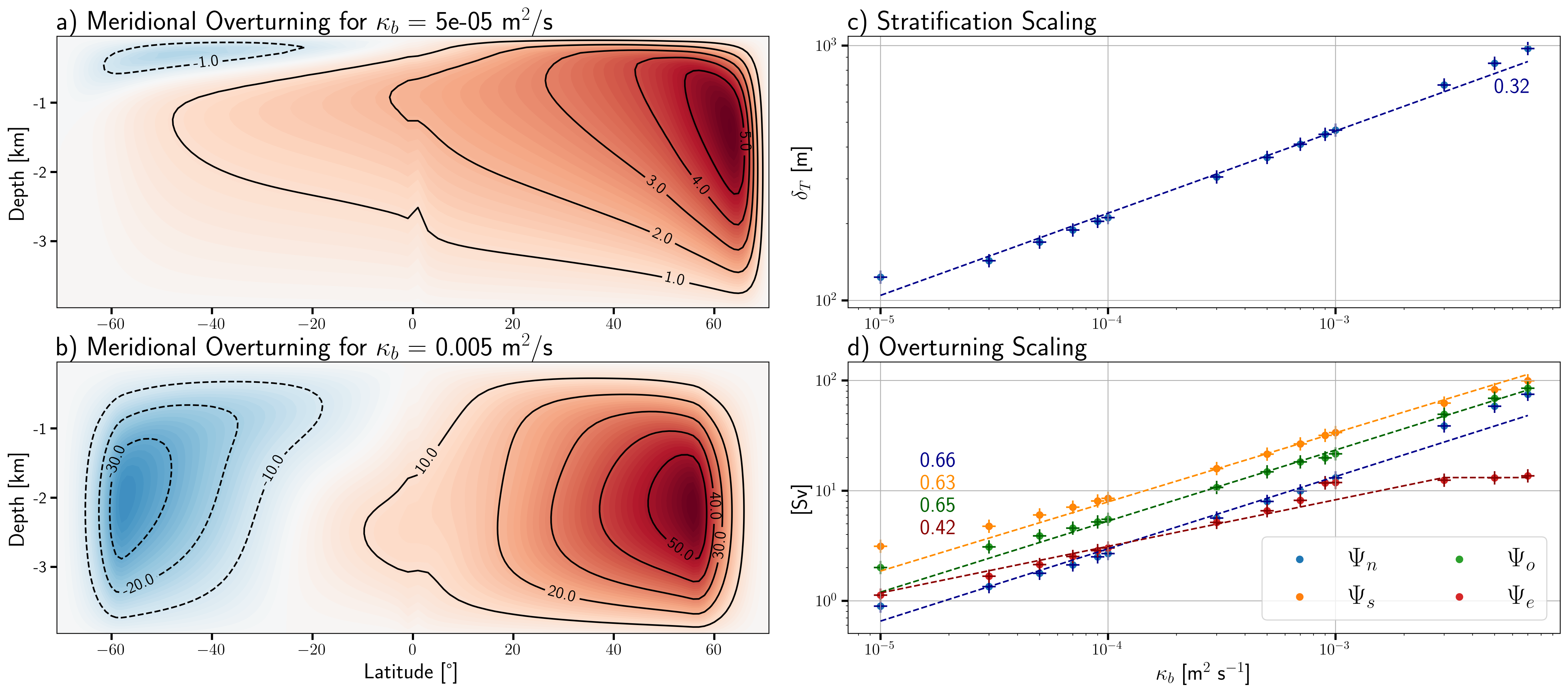}}	
		\caption{\em \fontsize{10.5}{12.5}\selectfont Steady state overturning streamfunction of the reference case for (a) $\kappa_b = 5\times10^{-5}$ m$^2$ s$^{-1}$ and (b) $\kappa_b = 5\times10^{-3}$ m$^2$ s$^{-1}$. (c) Scaling of the pycnocline depth, diagnosed as the depth of the tropical (averaged over $30$°S-$30$°N) $6$°C eastern boundary isotherm. (d) Scaling of NOC ($\Psi_n$), SOC ($\Psi_s$), symmetric overturning strength ($\Psi_o$) and cross-equatorial overturning strength ($\Psi_e$)  The dashed lines show the best-fit curve, and the numbers indicate the corresponding slopes of the corresponding curves.}
		\label{F:DH_scaling}
	\end{figure}

	Fig.~\ref{F:DH_scaling}a,b shows the overturning streamfunction for $\kappa_b = 5\times10^{-5}$ m$^2$ s$^{-1}$ and $\kappa_b = 5\times10^{-3}$ m$^2$ s$^{-1}$, respectively. Increasing $\kappa_b$ strengthens the volume transport of both the NOC and SOC, while shifting their maxima downward and equatorward. This corresponds to a deepening of the northward (southward) NOC (SOC) upper branch and a widening of the eastern boundary downwelling region \citep{callies2012simple}. Defining the relative difference between the NOC ($\Psi_n$) and SOC ($\Psi_s$) strengths as $2(\Psi_n - \Psi_s)/(\Psi_n + \Psi_s)$, we find that it decreases from about $1$ to $0.5$, indicating that higher vertical diffusivity reduces overturning asymmetry between the two hemispheres.
	
	Outside the eastern boundary downwelling regions, and under steady-state conditions, upward temperature advection along the eastern and western boundaries is balanced, to first order, by downward diffusive mixing \citep{welander1971discussion}. This balance implies that the vertical upwelling velocity scales as:
	\begin{equation}
		W=\frac{\kappa_b}{\delta_T},
		\label{W_scale_1}
	\end{equation}
	where $W$ is the vertical velocity scale and $\delta_T$ the thermocline depth scale. By continuity, an equal amount of downwelling must occur elsewhere. Fig.~\ref{F:DH_vel}d shows that this downwelling is concentrated along the eastern boundary, where stratification vanishes or weakens toward the surface. Because it is supplied by a geostrophically balanced zonal flow, a scaling for eastern boundary downwelling follows from combining the scaled forms of equations~(\ref{ui}) and~(\ref{conte}):
	\begin{equation}
		W=\frac{\alpha g\delta_T^2\Delta T}{2\Omega a^2 \Delta\lambda\Delta \theta_b}.
		\label{W_scale_2}
	\end{equation}
	Here $\Delta T$ represents the scale of temperature variations within the thermocline and $\Delta\theta_b$ the latitude extent of a single hemisphere (i.e. $70$°). Equating (\ref{W_scale_1}) to (\ref{W_scale_2}) gives the thermocline depth scaling:
	\begin{equation}
		\delta_T=\left(\frac{2\Omega a^2 \Delta \lambda \Delta\theta_b\kappa_b }{\alpha g \Delta T}\right)^{1/3} ,
		\label{pyc_scale}
	\end{equation}
	recovering the classical $\kappa_b^{1/3}$ scaling of the pycnocline depth. Fig.~\ref{F:DH_scaling}c shows that the RGGOCM reproduces this $1/3$ scaling law closely, indicating that the assumptions underlying its derivation rest on strong physical grounds.
	
	By continuity, the downward vertical motion must be supplied by a meridional volume flux of scale (equation~(\ref{psib})):
	\begin{equation}
		\Psi_o = 2 a^2 \Delta \lambda \Delta \theta_b W
		= \frac{\alpha g\Delta T}{\Omega}\delta_T^2 = 
		\left(\frac{4 a^4 \Delta \lambda^2 \Delta \theta_b^2 \kappa_b^2 \alpha g \Delta T}{\Omega}\right)^{1/3},
		\label{psi_scale}
	\end{equation}
	where the pre-factor $2$ accounts for upwelling along both the eastern and western boundaries. By equating~(\ref{W_scale_1}) and (\ref{W_scale_2}), scaling~(\ref{psi_scale}) implicitly assumes a perfect balance between upwelling and downwelling within each hemisphere. Under asymmetric forcing, however, eastern boundary downwelling intensifies in the cooler hemisphere (Fig.~\ref{F:DH_vel}d). Equation~(\ref{psi_scale}) therefore represents the total unidirectional meridional transport in each hemisphere required to compensate for the hemispherically integrated upwelling. Using parameter values from the reference experiment (Table~\ref{T:Tab1}), and correcting for the fact that upwelling is not uniform over the eastern boundary, we obtain $\Psi_o = 11$~Sv. This matches closely with the $10.5$~Sv of hemispheric upwelling diagnosed in the RGGOCM (Fig.~\ref{F:DH_MOC}a).
	
	Although equation~(\ref{psi_scale}) provides no information about the difference in NOC and SOC strength, these three quantities can be related given that vertical upwelling is approximately equal in both hemispheres \citep{marotzke2000dynamics}:
	\begin{subequations}
		\begin{equation}
			\label{Psio_a}
			\Psi_n=\Psi_o-\Psi_e,
		\end{equation}
		\begin{equation}
			\label{Psio_b}
			\Psi_s=\Psi_o+\Psi_e,
		\end{equation}
	\end{subequations}
	where $\Psi_e$ is the cross-equatorial transport strength. The quantities $\Psi_n$, $\Psi_s$, and $\Psi_e$ can be diagnosed from the model output. Equations~(\ref{Psio_a}) and (\ref{Psio_b}) yield consistent estimates of $\Psi_o$ across a range of $\kappa_b$ values (not shown), with $\Psi_o$ closely following the expected scaling $\Psi_o \sim \kappa_b^{2/3}$ (Fig.~\ref{F:DH_scaling}d). A similar scaling holds for $\Psi_n$ and $\Psi_s$. In contrast, $\Psi_e$ depends much more weakly on $\kappa_b$, with an exponent of $0.42$, and becomes nearly independent of it for $\kappa_b > 10^{-3}$ m$^2$ s$^{-1}$. These results indicate that the variations of $\Psi_n$ and $\Psi_s$ with increasing $\kappa_b$ primarily reflect changes in hemispheric upwelling strength rather than overturning asymmetry. They further suggest that overturning asymmetry decreases with higher $\kappa_b$, consistent with the behavior anticipated in Fig.~\ref{F:DH_scaling}a,b.
	
\begin{figure}
		\centering
		{\includegraphics[width=\textwidth]{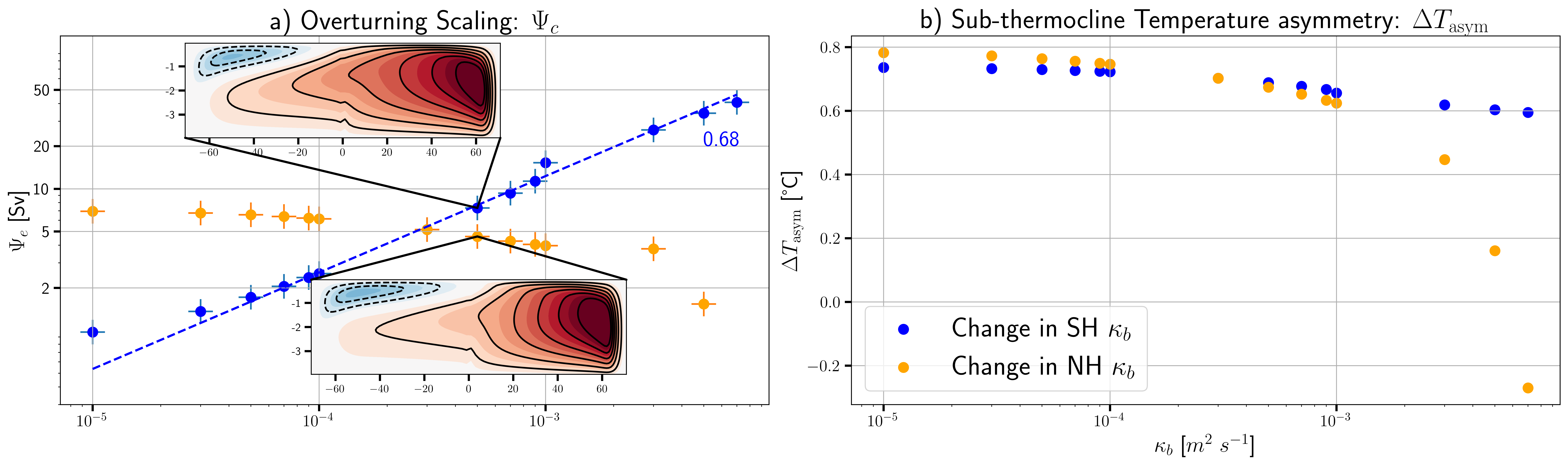}}	
		\caption{\em \fontsize{10.5}{12.5}\selectfont Scaling of (a) cross-equatorial overturning strength ($\Psi_e$) and (b) subthermocline temperature asymmetry ($\Delta T_{\mathrm{asym}}$) with vertical diffusivity $\kappa_b$, which is held constant at $3\times10^{-4}$ m$^{2}$ s$^{-1}$ in one hemisphere while varying in the opposite hemisphere. Dashed lines indicate best-fit curves. Insets in (a) show the corresponding overturning circulation for the highlighted data points, with contour intervals of $2.5$~Sv.}
		\label{F:DH_asym_scaling}
	\end{figure}

	To explore the reduced sensitivity of $\Psi_e$, we impose an equatorially asymmetric vertical diffusivity by increasing it in one hemisphere while fixing it at $3\times 10^{-4}$ m$^2$ s$^{-1}$ in the other. Fig.~\ref{F:DH_asym_scaling}a shows that enhancing SH diffusivity yields $\Psi_e \sim \kappa_{bS}^{2/3}$, while enhancing NH diffusivity reduces $\Psi_e$ without following a simple power law. This opposing response explains the weaker overall sensitivity of $\Psi_e$ in Fig.~\ref{F:DH_scaling}d.
	
	An increase in SH diffusivity, $\kappa_{bS}$, deepens the thermocline and weakens subthermocline stratification. To sustain cross-hemispheric transport under this condition, the subthermocline stratification along the western boundary must remain weaker than that along the eastern boundary (Fig.~\ref{F:DH_MOC}b). This is achieved through enhanced eastern boundary upwelling, supplied by the DWBC. The scaling of the subthermocline stratification depth thus follows the same advective–diffusive balance as in~(\ref{W_scale_1}), implying that the cross-equatorial transport naturally scales as $\Psi_e \sim \kappa_{bS}^{2/3}$, consistent with the results shown in Fig.~\ref{F:DH_asym_scaling}a.
	
	In contrast, increasing the NH diffusivity does not directly affect SH upwelling but alters the temperature range of SH subthermocline isotherms that sustain the SH subthermocline temperature asymmetry (Fig.~\ref{F:DH_MOC}b,c). We quantify this range as $\Delta T_{\mathrm{asym}} \equiv 1 - \min \{T_w(z,0^\circ)\}$. Fig.~\ref{F:DH_asym_scaling}b shows that increasing $\kappa_{bN}$ decreases $\Delta T_{\mathrm{asym}}$, whereas $\kappa_{bS}$ has little effect. For $\kappa_{bN} < 10^{-3}$ m$^2$ s$^{-1}$, $\Delta T_{\mathrm{asym}}$ is only weakly sensitive to diffusivity, but above this threshold it decreases sharply and eventually becomes negative, indicating the absence of SH subthermocline isotherms vanishes. This behavior is reflected in $\Psi_e$, which remains nearly insensitive to $\kappa_{bN}$ at low values but declines substantially once $\kappa_{bN} > 10^{-3}$ m$^2$ s$^{-1}$, disappearing entirely for $\kappa_{bN} = 9\times10^{-3}$ m$^2$ s$^{-1}$ (Fig.~\ref{F:DH_asym_scaling}a). The relative insensitivity of $\Psi_e$ for $\kappa_b > 10^{-3}$ observed in Fig.~\ref{F:DH_scaling}d therefore results from the counteracting effects of $\kappa_{bN}$ and $\kappa_{bS}$ at these diffusivities.
	
	The pronounced sensitivity of $\Delta T_{\mathrm{asym}}$ and $\Psi_e$ to $\kappa_{bN}$ suggests a possibly positive advective feedback: increased $\kappa_{bN}$ erodes the range of SH subthermocline isotherms, thereby reducing the DWBC’s southward advection of NH cold anomalies. While a detailed investigation of this mechanism is beyond the scope of this study, the main takeaway is that the overall sensitivity of $\Psi_e$ to $\kappa_b$ is reduced compared to the classical $2/3$ scaling, owing to two opposing effects: (i) in the SH, enhanced diffusivity strengthens eastern boundary upwelling and DWBC transport, whereas (ii) in the NH, enhanced diffusivity erodes the range of SH subthermocline isotherms that sustain the SH subthermocline asymmetry, thereby reducing $\Psi_e$.
	
	\section{Interhemispheric Flow with Adiabatic and Diffusive Upwelling}{\label{S:Channel}}
	In a purely diffusive setting, the RGGOCM sustains upwelling through vertical mixing. However, it is now well established that, in addition to vertical mixing, adiabatic wind-driven motions provide a crucial source of upwelling that helps close the interhemispheric overturning circulation \citep{marshall2012closure,talley2013closure}. In the Southern Ocean, the absence of meridional boundaries allows wind-driven motions to penetrate to great depths, influencing stratification and, consequently, the geostrophic flow within the basin \citep{toggweiler1995effect,wolfe2010sets,nikurashin2011theory}. By contrast, winds over enclosed basins are less effective, since meridional boundaries permit a shallow geostrophic return flow \citep{nikurashin2012theory}. In the following, we incorporate this adiabatic upwelling due to winds over the Southern Ocean into our framework.
	
	\subsection{Formulation and Model Domain}
\begin{figure}
		\captionsetup{justification=centering}
		\centering
		{\includegraphics[width=0.5\textwidth]{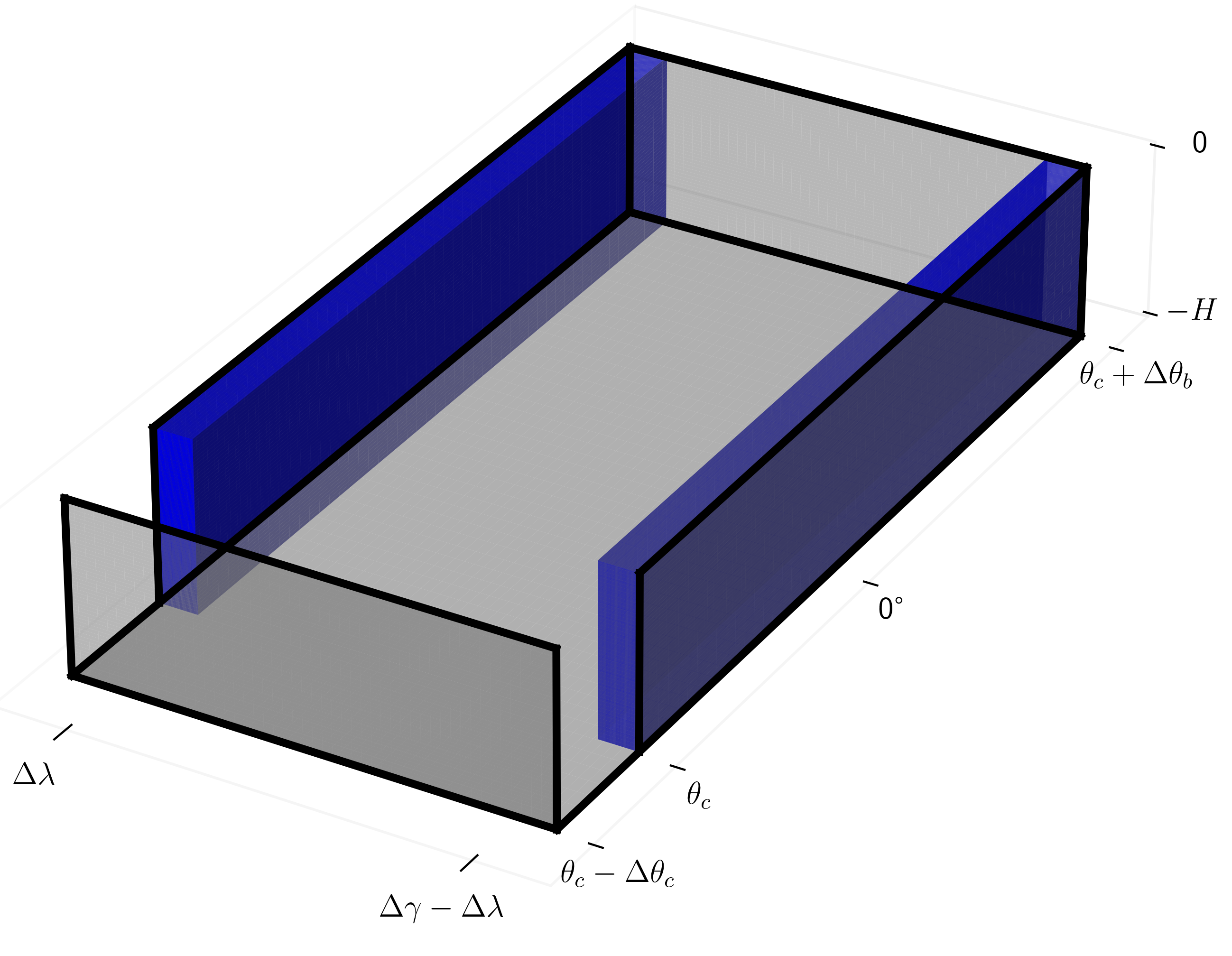}}
		\caption{\em \fontsize{10.5}{12.5}\selectfont Schematic of the model domain. As in Fig.~\ref{F:Fig1}, blue regions indicate the western and eastern boundary layers, each with zonal width $\Delta\lambda$. The channel extends from $\theta_c - \Delta\theta_c$ to $\theta_c$, while the basin extends from $\theta_c$ to $\theta_c + \Delta\theta_b$.}
		\label{F:Domain_Channel}
	\end{figure}

{\label{S:Channel_Domain}}
	We seek a reduced solution of the planetary geostrophic equations (\ref{xmom_bas})-(\ref{td_bas}) in the domain shown in Fig.~\ref{F:Domain_Channel}. The domain consists of a semi-enclosed basin extending from $-50$°S ($\theta_c$) to $70$°N ($\theta_n = \theta_c + \Delta \theta_b$) in latitude and from $0$°E to $60$°E ($\Delta\gamma$) in longitude. Its southern boundary connects to a zonally periodic re-entrant channel of the same longitudinal extent, spanning $-70$°S ($\theta_s=\theta_c - \Delta\theta_c$) to $-50$°S. The ocean bottom is assumed flat at a depth of 4~km ($H$). To represent an adiabatic upwelling pathway in the re-entrant channel, we include a zonal wind-stress forcing in equation~(\ref{xmom_bas}).
	
	The zonally periodic re-entrant channel can be regarded as an analogue of the Southern Ocean, while the semi-enclosed basin represents the Atlantic Ocean. In the current climate, most of the water upwelled adiabatically ultimately sinks in the high-latitude Atlantic \citep{cessi2019global}, whereas the formation of AABW is largely balanced by upwelling in the Indo-Pacific \citep{ferreira2018atlantic}. This distribution of upwelling and downwelling across multiple basins cannot be captured in our simplified configuration (Fig.~\ref{F:Domain_Channel}). Although the RGGOCM could, in principle, be extended to include multiple semi-enclosed basins, we leave this for future work.
	
	\subsection{Assumptions and Governing Equations}
	In the channel, the zonal uniformity of the surface forcing (wind stress and heat flux) motivates the assumption of no zonal variations in the solution. Consequently, the zonal pressure gradient term in equation~(\ref{xmom_bas}) makes a negligible contribution. Friction is also neglected, as its effect is small for $r \ll f$. Under these assumptions, the Eulerian flow in the re-entrant channel is purely wind-driven. The latitude–depth structure of this flow is described by the overturning streamfunction:
	\begin{equation}
		\overline{\psi}_c(\theta) = -a\cos(\theta)\Delta\gamma\frac{\tau_x(\theta)}{\rho_0 f},
		\label{psic_eul}
	\end{equation}
	where $\tau_x(\theta)$ is the zonal wind stress, $\Delta \gamma$ again the longitudinal width of the channel,  and $\rho_0 = 1025$ kg m$^{-3}$ is a reference density. Note that $\overline{\psi}_c$ depends only on latitude $\theta$, implying that the Eulerian vertical velocity is uniform with depth \citep{wolfe2011adiabatic}.
	
	For westerlies ($\tau_x > 0$), a positive overturning cell emerges in the channel, commonly referred to as the Deacon cell. This thermally indirect circulation steepens isopycnal slopes, increasing baroclinicity, but the associated mesoscale eddies act to flatten the isopycnals. The combination of these opposing effects defines the residual overturning circulation, which is the component of the flow that transports tracers. Its strength is given by
	\begin{equation}
		\psi_c^\dagger = \overline{\psi}_c + \psi^*,
		\label{psic_dagger}
	\end{equation}
	where $\psi_c^\dagger$ is the residual overturning streamfunction and $\psi^*$ is the eddy-induced streamfunction.
	
	The eddy-driven circulation is typically parameterized as being proportional to the isopycnal slope \citep{gent1990isopycnal}. However, this approach leads to singularities in the mixed layer. We avoid this by solving a boundary value problem \citep{ferrari2010boundary}:
	\begin{equation}
		\left(c_m^2\frac{d^2}{dz^2}-N^2\right)\psi^*_c=\alpha g\cos(\theta)\Delta\gamma K_{\mathrm{gm}}\frac{\partial T_c}{\partial \theta},
		\label{psic_edd}
	\end{equation}
	where $c_m^2$ is the squared baroclinic wave speed of mode $m$, $N^2$ is the Brunt–Väisälä frequency, and $K_{\mathrm{gm}}$ is the 
	(Gent-McWilliams) eddy diffusivity.
	
	Under the zonal-uniformity assumption, the residual latitude–depth circulation is fully determined by $\psi_c^\dagger$, with velocities given by
	\begin{align}
		-\frac{\partial \psi_c^\dagger}{\partial z} &= a\cos(\theta)\Delta\gamma v_c^\dagger,
		& \frac{\partial \psi_c^\dagger}{\partial \theta} &= a^2\cos(\theta) \Delta\gamma w_c^\dagger,
		\label{vcwc_res}
	\end{align}
	where $v_c^\dagger$ and $w_c^\dagger$ are the residual meridional and vertical velocities, respectively.
	
	Applying the same assumption to the thermodynamic equation~(\ref{td_bas}), the temperature evolution in the channel is governed by
	\begin{equation}
		\frac{\partial T_c}{\partial t} + \frac{v_c^\dagger}{a}\frac{\partial T_c}{\partial \theta} + w_c^\dagger \frac{\partial T_c}{\partial z}
		= \frac{\partial}{\partial z}\left(\kappa_c \frac{\partial T_c}{\partial z}\right) + c_c,
		\label{Tc}
	\end{equation}
	where $T_c$ is the channel temperature, $\kappa_c$ is the vertical diffusivity, and $c_c$ is the convective mixing tendency. In our boundary mixing formulation $\kappa_c \ll \kappa_b$. Moreover, we exclude meridional diffusion, which in Section 2 was introduced mainly for numerical stability. In flux-limited form, equation~(\ref{Tc}) can be solved stably without such a term (Appendix A).
	
	The assumptions and resulting equations for the semi-enclosed basin are described in Section~\ref{S:enclosed}\ref{S:Assumptions_enclosed_Domain} and equivalently implemented in the domain of Fig.~\ref{F:Domain_Channel}. The streamfunction over the full latitude extent of the domain is thus given by:
	\begin{align}
		\psi(z,\theta) =
		\begin{cases}
			\psi_c^\dagger & \text{for } \theta \leq \theta_c, \\
			\psi_b & \text{for } \theta > \theta_c,
		\end{cases}
		\label{psitotal}
	\end{align}
	
	\subsection{Boundary conditions}
	In the re-entrant channel, we impose a no-vertical-flux boundary condition for temperature at the ocean floor and a relaxation boundary condition at the ocean surface, as described by equation~(\ref{TopBC_T}).
	
	We use a zonal wind-stress profile:
	\begin{equation}
		\tau_x(\theta)=\tau_{\mathrm{max}}\sin\left(\frac{\pi}{2}\frac{\theta-\theta_s}{\theta_c-\theta_s}\right)\mathcal{H}(\theta_c-\theta),
		\label{taux}
	\end{equation}
	where $\tau_{\mathrm{max}}$ is the wind-stress amplitude near the interface and $\mathcal{H}$ is the Heaviside function.
	
	No-normal-flow conditions are applied at the ocean surface, the ocean floor, the southern end of the channel, and the northern end of the basin. In the channel, the first two conditions are satisfied by solving equation~(\ref{psic_edd}) subject to
	$\psi^\dagger_c(-H, \theta) = \psi^\dagger_c(0, \theta) = 0$. The procedure for satisfying these conditions in the basin is described in Section~\ref{S:enclosed}\ref{S:BC_enclosed_Domain}. The no-normal meridional flow condition in the channel is imposed by requiring $\partial_\theta T_c(z,\theta_s) = 0$, and $\tau_x(\theta_s) = 0$ by equation~(\ref{taux}). 
	
	Because the zonal mean basin temperature is mostly determined by the eastern boundary temperature (i.e. $\Delta\lambda \ll \Delta \gamma$), we impose
	\begin{equation}
		T_e(z,\theta_c)=T_c(z,\theta_c).
		\label{BC_int_1}
	\end{equation}
	For condition~(\ref{BC_int_1}) to yield a stable solution, we further require $\partial_t T_e \approx \partial_t T_c$ at $\theta = \theta_c$. Achieving exact equality would involve solving a complex nonlinear problem. In practice, stability is automatically ensured if
	\begin{equation}
		\frac{\partial T_e}{\partial \theta}(z,\theta_c) = 0.
		\label{BC_int_2}
	\end{equation}
	Condition~(\ref{BC_int_2}) enforces $w_e(z,\theta_c) = 0$, allowing $T_e(z,\theta_c)$ to evolve on similar timescales as $T_c(z,\theta_c)$, thereby ensuring that~(\ref{BC_int_1}) can be applied stably.
	
	Boundary conditions~(\ref{BC_int_1})-(\ref{BC_int_2}) close the problem for $T_e$ and $T_c$. A final boundary condition follows from a continuity of volume flux at the channel--basin interface:
	\begin{align}
		\frac{\Delta \gamma}{\Delta\lambda}\frac{\partial v_c^\dagger}{\partial z}
		&= \frac{\alpha g}{f^2+r^2}\left[\frac{f}{a\cos(\theta)\Delta\lambda}(T_e-T_w) - \frac{r}{2a}\frac{\partial}{\partial \theta}(T_e+T_w)\right],
		&& \forall z \text{ and }\theta=\theta_c, 
		\label{BC_int_3}
	\end{align}
	such that condition~(\ref{BC_int_3}) closes the problem for $T_w$. Note that boundary conditions~(\ref{BC_int_1})–(\ref{BC_int_3}) do not ensure a continuous advective temperature flux across the channel–basin interface. To account for this discontinuity, we define
	\begin{align*}
		\mathcal{E}(z)=v_c^\dagger \left[T_c-\frac{T_e+T_w}{2}\right], &&  \forall z \text{ and } \theta=\theta_c,
	\end{align*}
	which is added to the $T_e$ and $T_w$ tendency equations at the channel–basin interface. The term $\mathcal{E}$ represents an eddy temperature flux and is typically small compared to the total flux $v_c^\dagger T_c$.
	
	\subsection{Reference case}
	The model parameters used in the reference case, solved in the domain shown in Fig.~\ref{F:Domain_Channel}, are largely consistent with those listed in Table~\ref{T:Tab1}. Table~\ref{T:Tab2} lists the parameters that either differ from Table~\ref{T:Tab1} or are additionally required to solve equations~(\ref{psic_eul})–(\ref{Tc}). We choose $\kappa_b$ three times smaller compared to Section~\ref{S:enclosed}\ref{S:Sol_enclosed_Domain} to explore the role of adiabatic dynamics in the geostrophic GOC.
	
	\begin{table}[ht]
		\centering
		\begin{tabular}{l c c}
			\hline
			Parameter & Symbol & Value \\ 
			\hline
			Vertical diffusivity (boundary) & $\kappa_b$ & $1 \times 10^{-4} \ \mathrm{m^2} \ \mathrm{s^{-1}}$ \\
			Vertical diffusivity (channel) & $\kappa_c$ & $1 \times 10^{-7} \ \mathrm{m^2} \ \mathrm{s^{-1}}$\\
			Temperature asymmetry parameter & $T_n$ & $0.7$°C\\
			Minimal temperature of SH & $T_{\mathrm{min}}$ & $0$°C\\ 		
			Zonal wind-stress amplitude & $\tau_{\mathrm{max}}$& $0.2$ N m$^{-2}$\\
			Mesoscale diffusivity & $K_{gm}$ & $1\times10^3$ m$^{2}$ s$^{-1}$\\ 
			\hline
		\end{tabular}
		\caption{Model parameters used in reference run of semi-enclosed basin connected re-entrant channel (Fig.~\ref{F:Domain_Channel}).}
		\label{T:Tab2}
	\end{table}
	
\begin{figure}
		\captionsetup{justification=centering}
		\centering
		{\includegraphics[width=\textwidth]{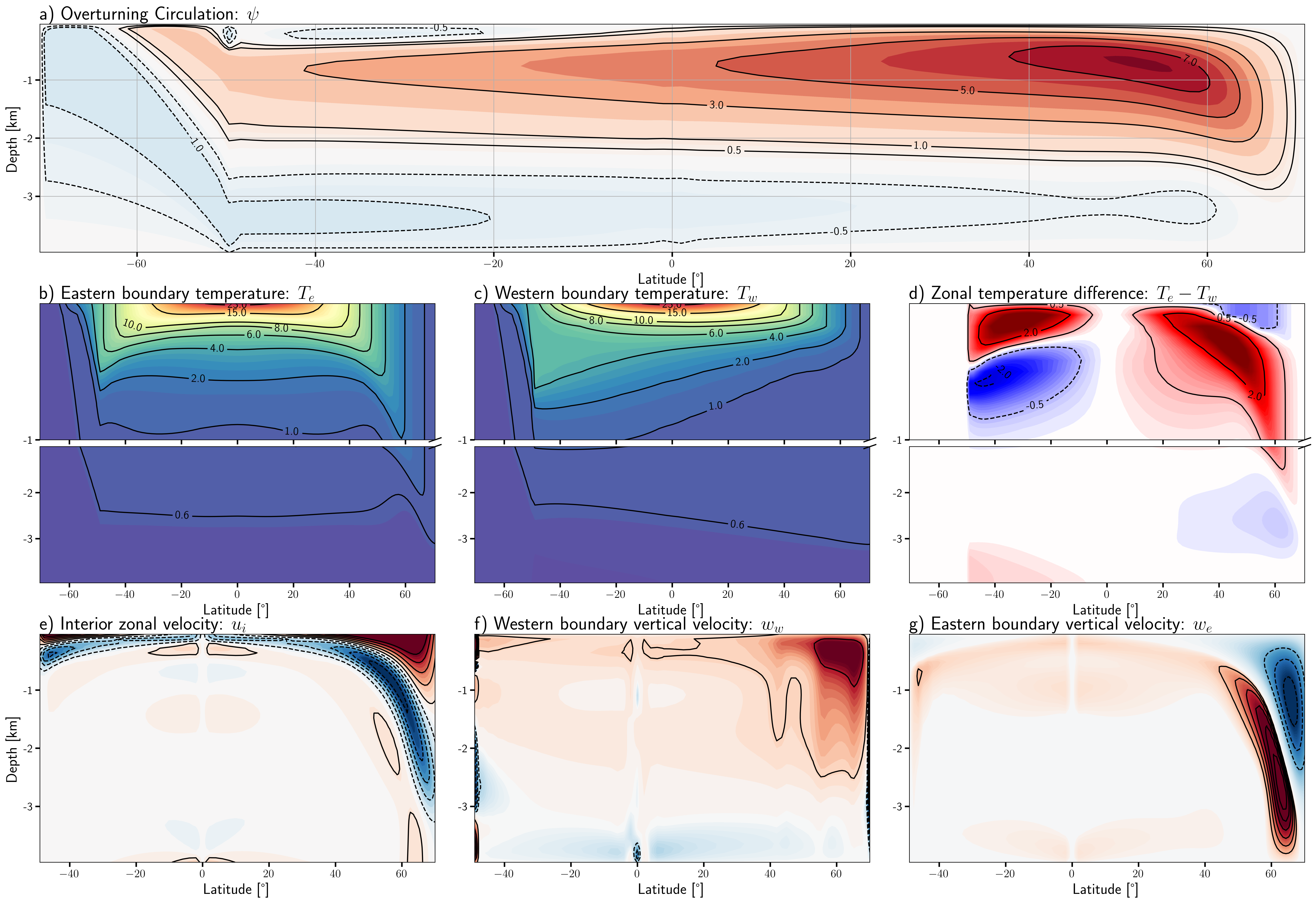}}
		\caption{\em \fontsize{10.5}{12.5}\selectfont Steady-state solution of the reference case with parameter values given in Table~\ref{T:Tab2}. (a) Overturning streamfunction (equation~(\ref{psitotal})). Temperature at the (b) eastern and (c) western boundary, and their (d) zonal difference. (e) Interior zonal velocity (contour interval: 0.2 cm s$^{-1}$). (f) Western boundary vertical velocity. (g) Eastern boundary vertical velocity. In panels (e)-(g) the contour intervals are equivalent to Fig.~\ref{F:DH_vel}. In panels (a), (e), (f), and (g), red shading denotes positive values and blue shading denotes negative values.}
		\label{F:Diagnostics_Channel}
	\end{figure}

	Fig.~\ref{F:Diagnostics_Channel} shows the steady-state solution of the reference case. The overturning streamfunction in the semi-enclosed basin (Fig.~\ref{F:Diagnostics_Channel}a) exhibits three distinct cells. The first is an interhemispheric cell, characterized by NH high-latitude sinking at a rate of $8.1$~Sv. The second is a shallow, negative overturning cell confined to the SH, with high-latitude sinking of $1$~Sv. These two cells are structurally analogous to the NOC and SOC described in Section~\ref{S:enclosed}\ref{S:Sol_enclosed_Domain} and will therefore be referred to as such. The third cell, not described in Section~\ref{S:enclosed}\ref{S:Sol_enclosed_Domain}, is an Abyssal Overturning Circulation (AOC), with $1.4$~Sv of uniformly distributed basin upwelling balanced by sinking in the channel.
	
	Of the $8.1$~Sv sinking in the NOC, about $5.5$~Sv is balanced by diffusive upwelling within the basin, while the remaining $2.6$~Sv returns adiabatically through the channel to the surface, where it is heated, flows northward, and re-enters the basin. The NOC is therefore sustained by a combination of adiabatic and diffusive processes. In contrast, all SOC upwelling occurs diffusively within the basin, making it diffusively controlled (i.e., vanishing in the limit $\kappa_b \to 0$). Similarly, the AOC is diffusively controlled, as all channel sinking is balanced by diffusive upwelling in the stratified abyssal basin (Fig.~\ref{F:Diagnostics_Channel}b,c).
	
	For discussion purposes, we separate the thermal structure of the solution into three categories: (1) thermocline isotherms, which outcrop on both sides of the equator within the basin; (2) subthermocline isotherms, which outcrop in the channel and in the high-latitude NH basin; and (3) abyssal isotherms, which outcrop only in the channel. In the channel, isotherms slope steeply downward toward the basin, with the slope determined by the balance between eddy- and wind-driven circulations. Within the basin, eastern boundary thermocline and subthermocline isotherms are nearly symmetric about the equator, relatively flat from the tropics to midlatitudes, and sharply outcrop at high latitudes (Fig.~\ref{F:Diagnostics_Channel}b). Western boundary thermocline isotherms are also relatively symmetric but gradually slope upward toward their outcrop location (Fig.~\ref{F:Diagnostics_Channel}c). In contrast, western boundary subthermocline isotherms are asymmetric about the equator, with a gradual upward slope toward the NH outcrop latitude. Eastern boundary abyssal isotherms remain relatively flat within the basin until sharply intersecting the ocean bottom or northern boundary, whereas western boundary abyssal isotherms gradually slope downward toward their boundary intersection point. This stratification produces a positive zonal temperature difference in the thermocline on both sides of the equator, and an anti-symmetric zonal temperature difference in the subthermocline and abyssal layers (Fig.~\ref{F:Diagnostics_Channel}d).
	
	The thermocline asymmetry in Fig.~\ref{F:Diagnostics_Channel}d closely resembles that in Fig.~\ref{F:DH_MOC}c. It is linked to eastern boundary upwelling overlying downwelling, forming a three-layer zonal circulation (Fig.~\ref{F:Diagnostics_Channel}e–g). Eastern boundary downwelling deepens the thermocline relative to the western boundary, where upwelling prevails (Fig.~\ref{F:Diagnostics_Channel}f). The asymmetry is inherently diffusive: the steady-state amplitudes of eastern boundary downwelling, the underlying upwelling, and western boundary upwelling depend on the mixing strength.
	
	The SH subthermocline asymmetry in Fig.~\ref{F:Diagnostics_Channel}c resembles that in Fig.~\ref{F:DH_MOC}c but is confined to shallower depths. The associated eastern boundary upwelling identified in Section~\ref{S:enclosed}\ref{S:Sol_enclosed_Domain} is also much weaker (Fig.~\ref{F:Diagnostics_Channel}g), reflecting the partly adiabatic nature of the asymmetry. Roughly $60$\% of the cross-equatorial NOC transport is sustained by adiabatic upwelling in the channel, with the remaining $40$\% arising diffusively through eastern SH boundary upwelling. Hence, the SH subthermocline asymmetry is primarily maintained by its adiabatic contribution. Specifically, subduction of relatively warm channel surface waters into the western boundary subthermocline deepens the isotherms there, producing the SH negative subthermocline asymmetry (Fig.~\ref{F:Diagnostics_Channel}g). Moreover, in the weak-mixing limit, the steady-state equations~(\ref{Te}) and~(\ref{Tw}) imply that the streamfunction remains nearly constant along subthermocline isotherms outside the NH outcropping region \citep{wolfe2011adiabatic}, where the meridionally flat $T_e$ enforces $u_i \approx 0$ (Fig.~\ref{F:Diagnostics_Channel}b,e). This near-constancy requires the SH adiabatic subthermocline asymmetry to reverse across the equator. Consequently, the NH positive subthermocline asymmetry (Fig.~\ref{F:Diagnostics_Channel}d) also acquires an adiabatic component, arising from the same streamfunction constraint along isotherms.
	
	The SOC is associated with the SH thermocline asymmetry, whereas the NOC is linked to the NH thermocline asymmetry and the subthermocline asymmetry in both hemispheres. Both cells gradually strengthen along their poleward path, sustained by diffusive thermocline upwelling at the eastern and western boundaries (Fig.~\ref{F:Diagnostics_Channel}f–g). As in Section~\ref{S:enclosed}\ref{S:Sol_enclosed_Domain}, the poleward flow turns eastward at the basin boundary, sinks along the weakly stratified eastern boundary, flows westward at depth (Fig.~\ref{F:Diagnostics_Channel}e) and rejoins the western boundary as a DWBC (Fig.~\ref{F:Diagnostics_Channel}a). Because the basin terminates at $50$°S, the SH experiences much weaker accumulated upwelling and a shallower meridional temperature gradient. In addition, the SOC lacks any adiabatic contribution. Together, these factors explain why the SOC remains much weaker and shallower than the NOC.
	
	The SH positive abyssal asymmetry originates from the input cold channel bottom waters into the western boundary (Fig.~\ref{F:Diagnostics_Channel}c), which are advected northward by a DWBC. An eastward bottom flow, partly fed by the DWBC, drives eastern boundary upwelling and western boundary downwelling (Fig.~\ref{F:Diagnostics_Channel}f,g). These tendencies shoal the abyssal thermocline depth along the eastern boundary and deepen it along the western boundary (Fig.~\ref{F:Diagnostics_Channel}b,c). In the SH, where $T_e > T_w$, the cold DWBC inflow sustains the asymmetry. At the equator, where $T_e = T_w$, this anomalous cold western boundary advection vanishes, and the vertical velocities act to establish $T_e < T_w$ in the NH. This marks the reversal of the abyssal asymmetry and the emergence of an interhemispheric AOC that gradually supplies eastern boundary upwelling sustaining the diffusive abyssal asymmetry.
	
	%As the current advects northward and reaches the equator, it induces an eastward flow that produces eastern boundary upwelling and western boundary downwelling (Fig.~\ref{F:Diagnostics_Channel}e–g), enabling cross-equatorial transport. Because the downwelling is stronger, eastern boundary upwelling is maintained primarily by the northward abyssal current. The resulting flow field and stratification support an interhemispheric negative overturning circulation, closely analogous to circulations generated by near-boundary mixing \citep{callies2018dynamics,drake2020abyssal}. Crucially, this steady-state circulation depends on eastern boundary upwelling and is therefore diffusive in nature. The supply of cold abyssal waters into the channel is thus limited by the extent of eastern boundary upwelling, and ultimately by the strength of diapycnal mixing \citep{nikurashin2011theory}.
	
	\begin{figure}
		\captionsetup{justification=centering}
		\centering
		{\includegraphics[width=0.5\textwidth]{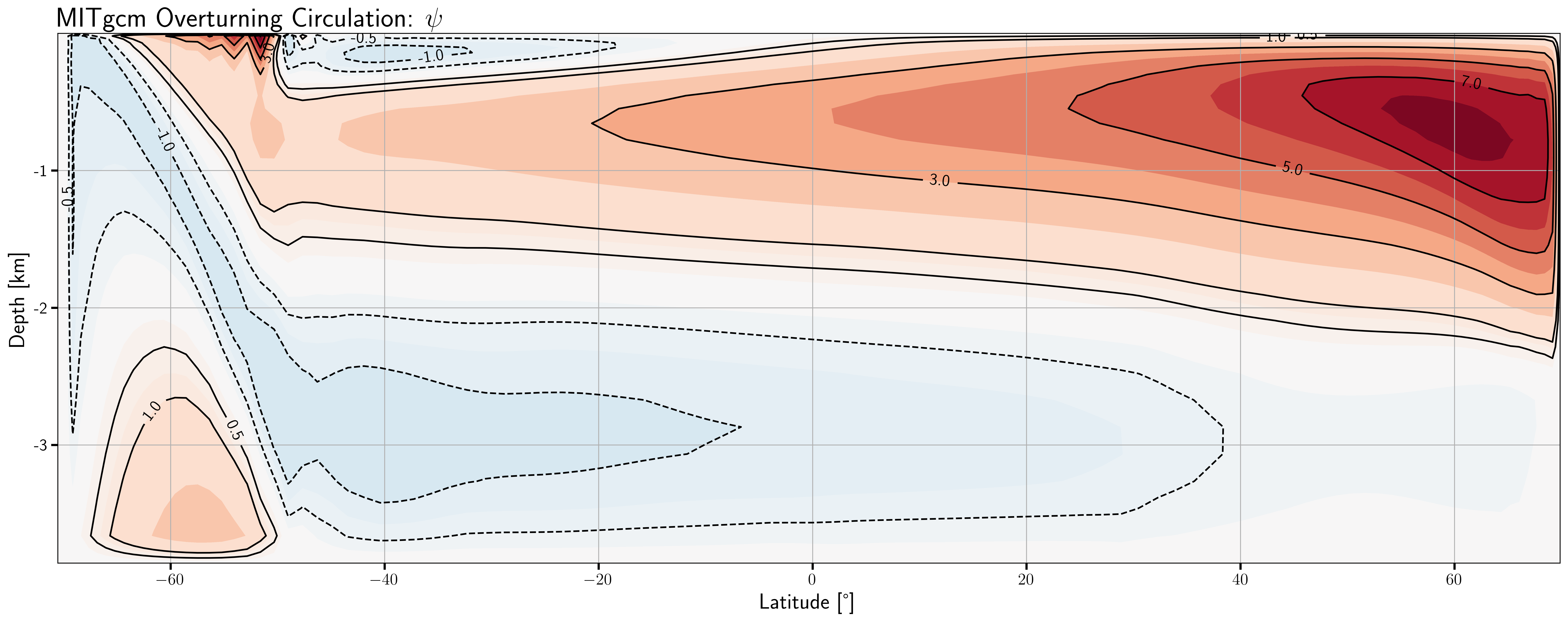}}
		\caption{\em \fontsize{10.5}{12.5}\selectfont Steady-state overturning streamfunction after 5000 years of integration with MITgcm (see Appendix B for experimental details).}
		\label{F:MITgcm_Channel}
	\end{figure}
	
	Fig.~\ref{F:MITgcm_Channel} shows the steady-state overturning streamfunction from a $5000$-year MITgcm integration in a domain equivalent to Fig.~\ref{F:Domain_Channel} (see Appendix B for details). Similar to Fig.~\ref{F:Diagnostics_Channel}a, the overturning exhibits three distinct cells: an interhemispheric mid-depth cell with a northern sinking rate of $8.4$~Sv, an abyssal cell with a maximum strength of $1.4$~Sv, and a shallow surface cell with a maximum strength of $1.1$~Sv. Of the mid-depth transport, about $2.8$~Sv upwells adiabatically in the channel. The associated stratification and velocity fields along the western and eastern boundaries (not shown) closely resemble Figures~\ref{F:Diagnostics_Channel}b–g. 
	
	We note the presence of an abyssal positive overturning cell in the channel (Fig.~\ref{F:MITgcm_Channel}). This cell arises from the absence of a ridge in the re-entrant channel, resulting in a  deep return flow \citep{nikurashin2012theory}. It is not reproduced in the RGGOCM due to the nature of the $\psi^\dagger_c$ vertical boundary condition, but can be regarded as dynamically irrelevant since it does not contribute to the heat transport.
	
	\subsection{Adiabatic Adjustment}\label{S:Channel_Adj}
\begin{figure}
		\captionsetup{justification=centering}
		\centering
		{\includegraphics[width=\textwidth]{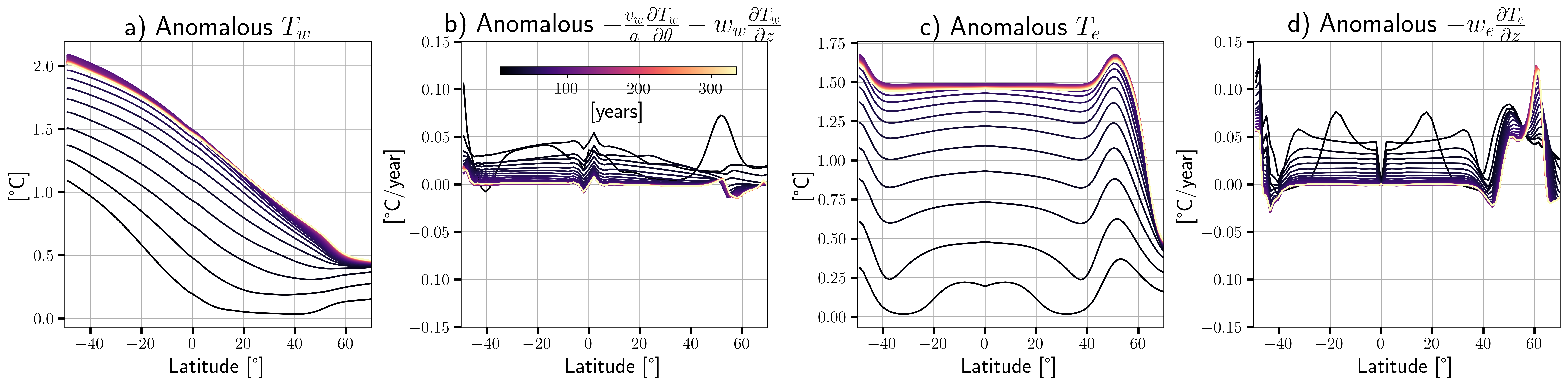}}
		\caption{\em \fontsize{10.5}{12.5}\selectfont Spin-up starting from a "reference-like" steady-state with $\kappa_b = 1\times10^{-7}$ m$^2$ s$^{-1}$ and $\tau_{\mathrm{max}}=0$ N m$^{-2}$, toward steady-state with $\tau_{\mathrm{max}}=0.2$ N m$^{-2}$. Latitude dependence of anomalous (a) $T_w$, (b) advective western boundary tendency, (c) $T_e$ and (d) advective eastern boundary tendency. All anomalies are calculated relative to the "reference-like" steady-state and averaged over a $200$-$1200$~m depth range.}
		\label{F:Spin_ad}
	\end{figure}

	In the previous section, it was suggested that the subthermocline asymmetry has a partly adiabatic origin. To illustrate this, we perform a spin-up experiment starting from a steady state of similar to the reference case, but with $\kappa_b = 1\times10^{-7}$ m$^2$ s$^{-1}$ and $\tau_{\mathrm{max}}=0$ N m$^{-2}$, after which $\tau_{\mathrm{max}}$ is increased to $0.2$ N m$^{-2}$. Strengthening the wind stress steepens isotherms in the channel and enhances the heat flux into the low-latitude subthermocline western boundary (Fig.~\ref{F:Spin_ad}a), generating a positive $T_w - T_e$ anomaly. This warm anomaly is advected northward by the NOC upper branch and transmitted to the eastern boundary at the equator, where it induces eastward surface flow and anomalous downwelling that raises and spreads the bell-shaped $T_e$ warm anomaly centered at the equator (Fig.~\ref{F:Spin_ad}b-d). The accompanying western boundary upwelling slows the northward propagation of the $T_w$ anomaly (Fig.~\ref{F:Spin_ad}b), ultimately reversing the $T_e - T_w$ sign across the equator and allowing the adiabatic NOC to cross.
	
	This mechanism parallels the spin-up process described in Section~\ref{S:enclosed}\ref{S:DH_Adjust}. However, in the adiabatic case, the system evolves toward a state in which the streamfunction remains constant along subthermocline isotherms. This is shown in Fig.~\ref{F:Spin_ad}b,d, where the advective fluxes at both boundaries decay to zero outside the convective regions and $T_e$ evolves to a meridionally flat profile (Fig.~\ref{F:Spin_ad}c). Hence, cross-diapycnal transport is no longer required to sustain subthermocline asymmetries (as in Section~\ref{S:enclosed}\ref{S:Sol_enclosed_Domain}); instead, these asymmetries persist as long as heat continues to advectively enter the western boundary from the channel.
	
	\subsection{Scaling of Stratification and Overturning Circulation}
	Similar to Section~\ref{S:enclosed}\ref{S:Scaling_DH}, we seek to establish scaling relations between key model parameters, the stratification, and the overturning strength. In the discussion above, we distinguished three stratification regimes: the thermocline, the subthermocline, and the abyssal thermocline. Each regime is characterized by a zonal temperature asymmetry arising from either adiabatic or diabatic processes. Instead of a single stratification depth scale as in Section~\ref{S:enclosed}\ref{S:Scaling_DH}, we now discuss three distinct depth scales: the thermocline depth ($\delta_T$), the subthermocline depth ($\delta_{ST}$), and the abyssal thermocline depth ($\delta_A$).
	
	To diagnose these scales, we calculate $\delta_T$ and $\delta_{ST}$ as the depths of the $6$°C and $3$°C tropical (average $30$°S–$30$°N) eastern boundary isotherm, respectively. The abyssal thermocline depth, $\delta_A$, is defined as the depth of the $0.7$°C western boundary  isotherm at the channel–basin interface ($50$°S). Since this isotherm outcrops only within the channel, it is by definition an abyssal isotherm.
	
	\subsubsection{Scaling with $\kappa_b$}
\begin{figure}
		\captionsetup{justification=centering}
		\centering
		{\includegraphics[width=\textwidth]{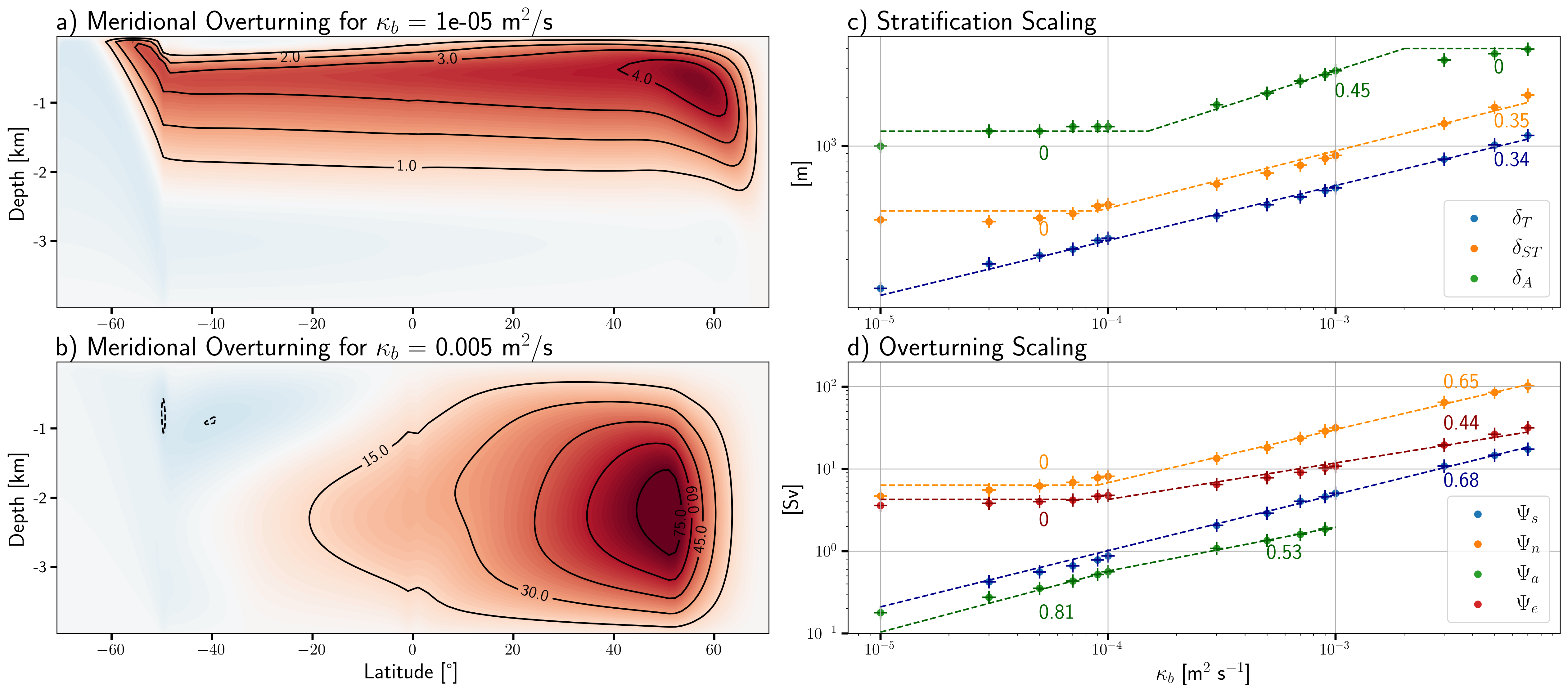}}
		\caption{\em \fontsize{10.5}{12.5}\selectfont Steady-state overturning streamfunction for (a) $\kappa_b = 1 \times 10^{-5}$ m$^2$ s$^{-1}$ and (b) $\kappa_b = 5 \times 10^{-3}$ m$^2$ s$^{-1}$. (c) Scaling of thermocline ($\delta_T$), subthermocline ($\delta_{ST}$) and abyssal thermocline ($\delta_A$). (d) Scaling of NOC ($\Psi_n$), SOC ($\Psi_s$), AOC ($\Psi_a$) and cross-equatorial ($\Psi_e$) flow strength. NOC (SOC) strength is computed as maximum positive (negative) value in the NH (SH above the $6$°C western boundary isotherm). The AOC strength is calculated as the mean overturning strength below the $0.8$°C isotherm. Values next to sloping lines give best fit slope values.}
		\label{F:Channel_kv_scaling}
	\end{figure}

	Fig.~\ref{F:Channel_kv_scaling}a,b show the steady-state overturning circulation for two values of $\kappa_b$. Reducing $\kappa_b$ to $10^{-5}$ m$^{2}$ s$^{-1}$, a factor 10 smaller than  in the reference case (Table~\ref{T:Tab2}),  weakens basin upwelling and lowers the NOC sinking rate to $5.2$~Sv. At the same time, adiabatic upwelling strengthens slightly, increasing its relative share of the NOC return flow. In contrast, increasing $\kappa_b$ to $5\times10^{-3}$ m$^{2}$ s$^{-1}$ enhances diffusive upwelling and raises the required NOC sinking to $72$~Sv. Under these conditions, the adiabatic pathway of the NOC disappears and instead contributes to SOC downwelling, which deepens and strengthens to $15$~Sv compared to the reference solution (Fig.~\ref{F:Diagnostics_Channel}a). Moreover, an overturning circulation analogous to the AOC in Fig.~\ref{F:Diagnostics_Channel}a and \ref{F:Channel_kv_scaling}a no longer exists.
	
	Fig.~\ref{F:Channel_kv_scaling}c shows how the three depth scales vary with $\kappa_b$. As discussed in Section~\ref{S:enclosed}\ref{S:Scaling_DH}, the thermocline depth scales as $\kappa_b^{1/3}$ (equation~(\ref{pyc_scale})), reflecting a balance between uniform diffusive upwelling and downwelling along the high-latitude eastern boundary, driven by convergent eastward geostrophic flow. The subthermocline depth follows the same scaling for $\kappa_b \gtrsim 10^{-4}$~m$^{2}$~s$^{-1}$, indicating that for sufficiently strong mixing, an analogous advective–diffusive balance governs the deeper layers. At smaller diffusivities, however, $\delta_{ST}$ asymptotes to a minimal value of about $350$~m and becomes independent of $\kappa_b$. In this weak-mixing regime, the subthermocline stratification is instead controlled by the adiabatic heat flux at the subthermocline western boundary-channel interface.
	
	The scaling of the NOC ($\Psi_n$), SOC ($\Psi_s$), AOC ($\Psi_a$), and cross-equatorial flow ($\Psi_e$) strengths is shown in Fig.~\ref{F:Channel_kv_scaling}d. As expected from equation~(\ref{psi_scale}), the SOC strength, governed by diffusive upwelling, scales as $\Psi_s \sim \kappa_b^{2/3}$. The NOC exhibits similar behavior for $\kappa_b \gtrsim 10^{-4}$~m$^{2}$~s$^{-1}$ but becomes independent of diffusivity at smaller values. This transition mirrors that of $\delta_{ST}$: in the weak-mixing regime, adiabatic dynamics determine $\delta_{ST}$ and thus the subthermocline asymmetry. As discussed in Section~\ref{S:Channel}\ref{S:Channel_Adj}, the adiabatic flow is characterized by the near constancy of the streamfunction along subthermocline western boundary isotherms. Consequently, the channel dynamics set the NOC strength in the low-mixing limit. Because these isotherms also cross the equator, the cross-equatorial transport $\Psi_e$ likewise becomes independent of $\kappa_b$ in this regime (Fig.~\ref{F:Channel_kv_scaling}d) and is therefore also governed by adiabatic channel dynamics.
	
	For $\kappa_b \gtrsim 10^{-4}$~m$^{2}$ s$^{-1}$, $\Psi_e$ scales as $\kappa_b^{0.45}$, consistent with Fig.~\ref{F:DH_asym_scaling}d. The reduced sensitivity of $\Psi_e$ to $\kappa_b$, relative to $\Psi_n$ and $\Psi_s$, was discussed in Section~\ref{S:enclosed}\ref{S:Scaling_DH} and implies that hemispheric overturning asymmetries weaken as $\kappa_b$ increases. Furthermore, higher basin diffusivity steepens the isotherm slope in the channel, thereby reducing adiabatic upwelling in the NOC. As shown in Fig.~\ref{F:Channel_kv_scaling}b, this upwelling pathway eventually disappears altogether, such that all cross-equatorial flow is upwelled along the eastern boundary, as in Section~\ref{S:enclosed}\ref{S:Sol_enclosed_Domain}.
	
	%At large $\kappa_b$, by contrast, cross-equatorial adiabatic flow disappears (Fig.~\ref{F:Channel_kv_scaling}b), because the isotherms slopes in the channel are so large that the channel overturning streamfunction is negative everywhere. In this regime, cross-equatorial flow is entirely controlled by diffusive processes and the associated SH eastern boundary upwelling, as for the overturning structure described in Section~\ref{S:enclosed}\ref{S:Sol_enclosed_Domain}.
	
	To understand the scaling of $\delta_A$ and $\Psi_a$ with $\kappa_b$, we first note that, unlike the NOC and SOC—where high-latitude sinking results from convergent eastward geostrophic flow—the AOC sinking arises from adiabatic channel dynamics. In this case, eastern boundary diffusive upwelling balances adiabatic channel downwelling, yielding the scaled equality \citep{ito2008control}, obtained from the equations and (\ref{psic_eul}) and (\ref{psic_edd}) :
	\begin{align}
		\Psi_a &= a\Delta\gamma\left(\frac{K_{gm} \delta_A}{a\Delta\theta_c}-\frac{\tau_{\mathrm{max}}}{2\Omega\rho_0}\right)
		= a^2\Delta\lambda\Delta\theta_b \frac{\kappa_b}{\delta_A}.
		\label{Psia_1}
	\end{align}
	Here,  we have neglected the nonlocal contribution of the eddy-driven flow (equation~(\ref{psic_edd})), a reasonable approximation if the abyss is well stratified. 
	
	Equation (\ref{Psia_1}) leads to a quadratic equation for $\delta_A$, where the negative root corresponds to the physically relevant solution:
	\begin{align}
		\delta_A &= -\frac{\overline{\Psi}\Delta\theta_c}{2K_{gm} \Delta\gamma}\left(-1-\sqrt{1+\phi_a}\right),
		&& \text{with } \phi_a=\frac{4K_{gm} \kappa_b a^2\Delta\lambda\Delta\gamma\Delta\theta_b}{\Delta\theta_c\overline{\Psi}^2},
		\label{pyc_a_solution}
	\end{align}
	and $\overline{\Psi}=a\Delta\gamma\tau_{\mathrm{max}}/(2\Omega\rho_0)$. Reinserting (\ref{pyc_a_solution}) into (\ref{Psia_1}), we obtain:
	\begin{align}
		\Psi_a=-\frac{\overline{\Psi}}{2}\left(1-\sqrt{1+\phi_a}\right).
		\label{psi_a_solution}	
	\end{align}
	We now analyze (\ref{pyc_a_solution})–(\ref{psi_a_solution}) in two limiting cases \citep{nikurashin2011theory}. First we consider a limit corresponding to a wind-dominated regime, where $\phi_a\ll 1$, such that $\sqrt{1+\phi_a}\approx 1+\phi_a/2$. Inserting this into (\ref{pyc_a_solution}) and (\ref{psi_a_solution}), we have:
	\begin{equation}
		\begin{aligned}
			\delta_A &= \frac{\overline{\Psi}\,\Delta\theta_c}{K_{gm}\,\Delta\gamma}, \qquad
			\Psi_a &= \frac{\overline{\Psi}\,\phi_a}{4}.
		\end{aligned}
		\label{phia_limit1}
	\end{equation}
	In this regime, channel wind-driven dynamics dominate over abyssal basin mixing, so $\delta_A$ is independent of $\kappa_b$, while $\Psi_a$ scales linearly with $\kappa_b$. Fig.~\ref{F:Channel_kv_scaling}c–d confirms this behavior for $\kappa_b \lesssim 10^{-4}$ m$^2$ s$^{-1}$, corresponding to a low-mixing regime. The model determined sensitivity of $\Psi_a$ ($\kappa_b^{0.8}$) is slightly weaker than the theoretical expectation, likely due to non-neglible contributions of $\xi_b$. 
	
	The second limit corresponds to the regime in which diffusive abyssal basin dynamics dominate over wind-driven motions in the channel. In this limit, $\phi_a \gg 1$, yielding:
	\begin{equation}
		\begin{aligned}
			\delta_A &= \frac{\overline{\Psi}\Delta\theta_c\sqrt{\phi_a}}{2K_{gm}\Delta\gamma}, \qquad
			\Psi_a  &= \frac{\overline{\Psi}\sqrt{\phi_a}}{2}.
		\end{aligned}
		\label{phia_limit2}
	\end{equation}
	Thus, when abyssal mixing dominates the channel dynamics, both $\delta_A$ and $\Psi_a$ scale as $\kappa_b^{1/2}$. This scaling is confirmed in Fig.~\ref{F:Channel_kv_scaling}c–d for $\kappa_b \gtrsim 10^{-4}$ m$^2$ s$^{-1}$. For even stronger mixing ($\kappa_b > 10^{-3}$ m$^2$ s$^{-1}$), however, $\delta_A$ becomes constant ($\kappa_b^0$) and equal to $H$, indicating that abyssal isotherms vanish from the basin as a result of the intense downward buoyancy transfer. Since the AOC is sustained by these abyssal isotherms, $\Psi_a$ is no longer well defined. Instead, the AOC merges with the SOC, forming a single negative overturning cell characterized by diffusive upwelling in the basin, and downwelling that is geostrophic within the basin and adiabatic within the channel (Fig.~\ref{F:Channel_kv_scaling}b).
	
	\subsubsection{Scaling with $\tau_{\mathrm{max}}$}
\begin{figure}
		\captionsetup{justification=centering}
		\centering
		{\includegraphics[width=\textwidth]{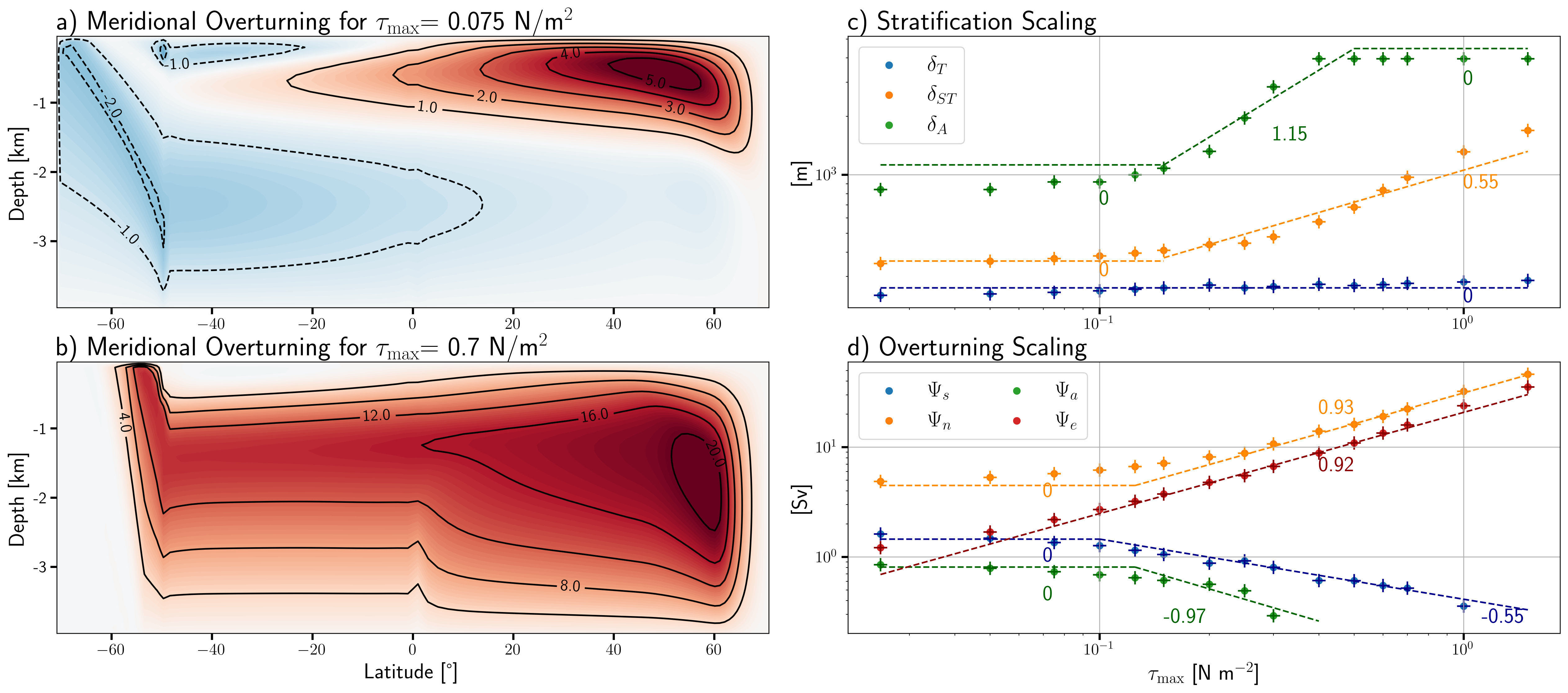}}
		\caption{\em \fontsize{10.5}{12.5}\selectfont Steady-state overturning streamfunction for (a) $\tau_{\mathrm{max}} = 0.075$ N m$^{-2}$ and (b) $\tau_{\mathrm{max}} = 0.7$ N m$^{-2}$. Panels (c)-(d) are analogues to Fig.~\ref{F:Channel_kv_scaling}c-d but now show scaling for $\tau_{\mathrm{max}}$.}
		\label{F:Channel_tau_scaling}
	\end{figure}

	Figures~\ref{F:Channel_tau_scaling}a–b show the steady-state overturning streamfunction for two values of $\tau_{\mathrm{max}}$. Reducing $\tau_{\mathrm{max}}$ to $0.075$ N m$^{-2}$ weakens and shallows the NOC, while strengthening and deepening the SOC and AOC. The NOC’s adiabatic upwelling pathway is notably reduced compared to Fig.~\ref{F:Diagnostics_Channel}a. In contrast, increasing $\tau_{\mathrm{max}}$ to $0.7$ N m$^{-2}$ strengthens, deepens, and shifts the NOC sinking northward, while the SOC shallows and weakens relative to Fig.~\ref{F:Diagnostics_Channel}a. The AOC disappears entirely, indicating the absence of abyssal isotherms in the basin.
	
	Since the $6$°C isotherm does not outcrop in the channel, changes in wind-forcing strength are not expected to affect $\delta_T$. Fig.~\ref{F:Channel_tau_scaling}c confirms this, showing only a very weak dependence of $\delta_T$ on $\tau_{\mathrm{max}}$.
	
	To examine how the subthermocline stratification and associated NOC strength vary with $\tau_{\mathrm{max}}$, we consider the regime where adiabatic channel dynamics dominate over basin diffusive upwelling. In this limit, all water sinking in the NH high latitudes is upwelled in the channel, yielding the scaled relation \citep{nikurashin2012theory}:
	\begin{equation}
		\Psi_n=\overline{\Psi}-K_{gm}\frac{\delta_{ST}\Delta\gamma}{\Delta\theta_c}=\frac{\alpha g\Delta T_{st}}{2\Omega}\delta_{ST}^2.
		\label{Psin_1}
	\end{equation}
	Here, $\Delta T_{st}$ represents the range of basin subthermocline isotherms. Equation~(\ref{Psin_1}) is quadratic in $\delta_{ST}$, with the positive solution:
	\begin{align}
		\delta_{ST}=\frac{\Omega K_{gm} \Delta\gamma}{\alpha g\Delta T_{st}\Delta \theta_c}\left(-1+\sqrt{1+\phi_{st}}\right), && \text{with } \phi_{st} = \frac{2\alpha g \Delta T_{st}\Delta\theta_c^2\overline{\Psi}}{\Omega K_{gm}^2 \Delta\gamma^2}.
		\label{pyc_st_solution}
	\end{align}
	Reinserting this solution into equation~(\ref{Psin_1}) gives:
	\begin{equation}
		\Psi_n=\frac{\Omega K_{gm}^2\Delta\gamma^2}{2\alpha g \Delta T_{st} \Delta\theta_c^2}\left(-1+\sqrt{1+\phi_{st}}\right)^2.
		\label{psi_st_solution}
	\end{equation}
	As with solution~(\ref{pyc_a_solution}), we now examine limiting behaviors. When wind-driven  dynamics dominate over eddy-driven 
	dynamics in the channel, $\phi_{st} \gg 1$, and the solutions reduce to:
	\begin{equation}
		\begin{aligned}
			\delta_{ST} &= \sqrt{\frac{2\overline{\Psi}\Omega}{\alpha g \Delta T_{st}}}, \qquad
			\Psi_n &= \overline{\Psi}.
		\end{aligned}
		\label{phist_limit1}
	\end{equation}
	Hence, in this limit, $\delta_{ST} \sim \tau_{\mathrm{max}}^{1/2}$ and $\Psi_n \sim \tau_{\mathrm{max}}^{1}$, consistent with Fig.~\ref{F:Channel_tau_scaling}c,d for $\tau_{\mathrm{max}} \gtrsim 0.1$~N~m$^{-2}$. This demonstrates that, under strong wind forcing, NOC sinking primarily acts to compensate for wind-driven channel upwelling \citep{wolfe2011adiabatic}.
	
	In this regime, $\Psi_s$ roughly scales as $\tau_{\mathrm{max}}^{-1/2}$ (Fig.~\ref{F:Channel_tau_scaling}d), although its sensitivity varies with $\tau_{\mathrm{max}}$. This behavior arises because $\delta_{ST} \sim \tau_{\mathrm{max}}^{1/2}$, so stronger wind stress weakens the stratification below the thermocline. From the advective–diffusive balance (equation~(\ref{W_scale_1})), it then follows that $\Psi_s \sim \tau_{\mathrm{max}}^{-1/2}$.
	
	In the wind-driven limit, the cross-equatorial flow scales as $\Psi_e \sim \tau_{\mathrm{max}}^{1}$ (Fig.~\ref{F:Channel_tau_scaling}d), consistent with $\Psi_n$. As discussed earlier, this reflects the near constancy of the streamfunction along subthermocline isotherms in the adiabatic limit. Because $\Psi_s \sim \tau_{\mathrm{max}}^{-1/2}$, whereas $\Psi_n$ and $\Psi_e \sim \tau_{\mathrm{max}}^{1}$, the hemispheric overturning asymmetry strengthens with increasing $\tau_{\mathrm{max}}$, suggesting that the interhemispheric circulation is most effectively driven by wind-induced channel upwelling.
	
	The second limit of equation~(\ref{pyc_st_solution}) corresponds to a regime where eddy and wind effects nearly compensate, yielding a residual circulation close to zero ($\phi_{st} \ll 1$). In this case, equation~(\ref{Psin_1}) is no longer strictly valid, as adiabatic channel upwelling may not dominate over basin diffusive upwelling. For sufficiently small $\phi_{st}$, $\Psi_n$ is thus controlled primarily by diffusive processes rather than channel dynamics. Figures~\ref{F:Channel_tau_scaling}c,d confirm this: at low $\tau_{\mathrm{max}}$, both $\delta_{ST}$ (and consequently $\Psi_s$) and $\Psi_n$ become independent of $\tau_{\mathrm{max}}$, consistent with an advective–diffusive balance.
	
	Finally, revisiting the two limiting values of $\phi_{a}$ (equations~(\ref{phia_limit1})–(\ref{phia_limit2})), we find that when diffusion dominates ($\phi_a \gg 1$), $\delta_A$ and $\Psi_a$ become independent of $\tau_{\mathrm{max}}$. In contrast, for $\tau_{\mathrm{max}} \gtrsim 0.1$ N m$^{-2}$, i.e., in the wind-dominant limit ($\phi_a \ll 1$), $\delta_A$ and $\Psi_a$ approximately scale as $\tau_{\mathrm{max}}^1$ and $\tau_{\mathrm{max}}^{-1}$, respectively (Fig.~\ref{F:Channel_tau_scaling}c,d). These results agree with theoretical predictions. However, when $\tau_{\mathrm{max}} \gtrsim 0.4$ N m$^{-2}$, the AOC vanishes, as the channel isotherms steepen to the point where abyssal isotherms no longer extend into the basin.
	\section{Summary and Discussion}\label{S:SumDis}
	In this paper, we developed the RGGOCM, a reduced-dimensional model to understand to spatial structure of the three-dimensional GOC, in 
	particular its interhemispheric flow. 
	Earlier reduced models of the GOC ignored two key observational constraints: the overturning circulation beneath the Ekman layer is largely geostrophic \citep{johns2005estimating,frajka2019atlantic}, and diapycnal mixing, with associated cross-isopycnal transport, are concentrated near ocean boundaries \citep{polzin1997spatial,st2012turbulence}. Their omission is unsurprising, as faithfully representing these processes typically requires resolving the full three-dimensional dynamics.
	
	Using a three-dimensional ocean model with mixing confined to vertical boundaries, \cite{marotzke1997boundary} showed that vertical velocities peak near those boundaries. Although later studies emphasize that boundary-intensified mixing primarily drives diapycnal upwelling along sloping, rather than vertical, walls \citep{ferrari2016turning,mcdougall2017abyssal}, the simplified framework of \cite{marotzke1997boundary} remains a useful idealization of enhanced diapycnal velocities near ocean boundaries. In this boundary-mixing limit, the ocean interior is characterized by zonally flat isopycnals connecting to the eastern boundary. Building on this framework, \cite{callies2012simple} proposed a two-plane model of the GOC in a single hemisphere, in which the prognostic evolution of eastern and western boundary temperatures suffices to reconstruct the overturning circulation.
	
	In the first part of our study, we extend the \cite{callies2012simple} model to a double-hemisphere configuration. The governing equations remain largely unchanged, but with two key modifications. First, we add a frictional term to the momentum equations to avoid singular behavior at the equator. Second, we include a parameterized representation of pressure-gradient adjustment by Kelvin wave propagation across the equator \citep{kawase1987establishment,johnson2002theory}.
	
	We find that under weakly asymmetric surface temperature forcing, the RGGOCM produces a strongly asymmetric overturning circulation. The dominant cell in the cooler NH (referred to as the NOC) develops a pronounced cross-equatorial component, associated with a negative zonal subthermocline temperature difference in the SH. The southward DWBC induces upwelling along the SH eastern boundary, locally shoaling the mixed layer and strengthening the stratification below the thermocline. The resulting shallower mixed layer weakens the surface eastward flow, thereby reducing western boundary upwelling relative to the NH and leading to weaker stratification in the SH west. This stratification asymmetry between the eastern and western boundaries—maintained by anomalous upwelling and downwelling in the SH—sustains the subthermocline temperature contrast. These dynamics closely resemble those found in three-dimensional GCM studies \citep{klinger1999behavior,marotzke2000dynamics}.
	
	To understand how the RGGOCM attains its strongly asymmetric state, we conducted a spin-up simulation from symmetric initial conditions. The results show that Kelvin waves play a central role in transmitting NH cold anomalies from the western to the eastern boundary. This transmission induces upwelling in the east and downwelling in the west, warming the western boundary relative to the east. The ensuing reversal of the subthermocline temperature gradient allows the DWBC to cross the equator and maintain the negative subthermocline asymmetry through continued eastern boundary upwelling. The adjustment mechanism in the RGGOCM thus involves both advective and wave-mediated processes—consistent with the cross-equatorial adjustment pathways identified in previous studies \citep{kawase1987establishment,marotzke2000dynamics,johnson2002theory}. We therefore conclude that the RGGOCM captures the essential physics of the pressure-gradient reversal in three-dimensional models.
	
	In many earlier studies, scaling laws for the vertical diffusivity ($\kappa_b$) were derived under assumptions linking the zonal and meridional flows \citep{welander1971discussion,kuhlbrodt2007driving}. Owing to its simplicity, the RGGOCM enables a transparent derivation of the previously proposed $\kappa_b^{2/3}\Delta T^{1/3}$ scaling for the domain-integrated upwelling ($\Psi_o$), where $\Delta T$ characterizes the temperature contrast within the thermocline, without invoking such assumptions. The model further yields comparable scaling relations for the dominant NOC ($\Psi_n$) and the weaker SOC confined to the SH ($\Psi_s$). In contrast, the cross-equatorial flow exhibits a weaker dependence, scaling as $\Psi_e \sim \kappa_b^{0.4}$, implying that the hemispheric asymmetry of the overturning circulation diminishes as $\kappa_b$ increases. This reduction in asymmetry arises because enhanced vertical mixing erodes anomalously cold NH water before it can reach the equator via southward DWBC transport.
	
	The second part of our study was motivated by suggestions that the mid-depth circulation is largely adiabatically upwelled in the Southern Ocean \citep{lumpkin2007global,marshall2012closure,cessi2019global}. To represent this, we extended the double-hemispheric RGGOCM by adding a zonally periodic re-entrant channel forced by surface westerlies. For realistic parameters, the model produces a GOC consistent with three-dimensional MITgcm results and other studies using similar forcings and geometries \citep{wolfe2011adiabatic,nikurashin2012theory,jansen2018transient}. Specifically, it reproduces a mid-depth overturning cell sustained by adiabatic upwelling in the channel and diffusive upwelling in the basin, and an abyssal cell that upwells diffusively in the basin and sinks adiabatically in the channel—here referred to as the NOC and AOC, respectively. The model also exhibits a shallow diffusive overturning cell, analogous to the SOC in the fully enclosed basin set-up, which has received comparatively little attention.
	
	The RGGOCM provides insight into the geostrophic nature of the GOC. In particular, the adiabatic component of the NOC is linked to subthermocline temperature asymmetries. Unlike in the purely diffusive scenario (Section~\ref{S:enclosed}\ref{S:Sol_enclosed_Domain}), where such asymmetries are maintained by anomalous vertical (diapycnal) fluxes, in the adiabatic system they are sustained by a heat influx from the channel into the western boundary. This input generates a negative zonal temperature difference in the SH, driving the SH NOC. The asymmetry reverses sign across the equator through the adjustment described in Section~\ref{S:Channel}\ref{S:Channel_Adj}, analogous to the adjustment of a diffusive overturning circulation (Section~\ref{S:enclosed}\ref{S:DH_Adjust}). However, in the adiabatic regime, the circulation adjusts toward a state in which the streamfunction remains constant along subthermocline isotherms outside outcropping regions, with diapycnal fluxes confined to the outcropping latitudes \citep{wolfe2011adiabatic}.
	
	Unlike the subthermocline asymmetries, zonal temperature differences in the thermocline and abyss are sustained by purely diabatic processes. In the thermocline, anomalous eastern boundary downwelling deepens the mixed layer relative to the western boundary, where upwelling dominates. The resulting positive temperature asymmetry on both sides of the equator strengthens the NH NOC and sustains the weaker SH SOC \citep{marotzke1997boundary}. The abyssal asymmetry, by contrast, is maintained by diffusive eastern boundary upwelling supplied by a bottom northward DWBC, giving rise to an interhemispheric AOC. Studies employing more elaborate boundary-intensified mixing schemes with sloping topography \citep{callies2018dynamics,drake2020abyssal} likewise identify diapycnal eastern boundary upwelling, fed by a DWBC, as the mechanism sustaining abyssal temperature asymmetries.
	
	The thermocline depth is set purely by diffusive dynamics, scaling as $\delta_T \sim \kappa_b^{1/3}\Delta T^{-1/3}$, so that the SOC scales as $\Psi_s \sim \kappa_b^{2/3}\Delta T^{1/3}$. In contrast, the subthermocline stratification, $\delta_{ST}$, and the associated NOC, $\Psi_n$, follow distinct scaling relations in two regimes. In the adiabatic, wind-driven limit, the subthermocline stratification is controlled by channel dynamics. In this weak-mixing regime, the streamfunction remains constant along subthermocline isotherms, implying that all water upwelled in the channel must downwell in the NH. These arguments yield scalings in which $\delta_{ST}$ and $\Psi_n$ become independent of $\kappa_b$, instead scaling as $\delta_{ST} \sim \tau_{\mathrm{max}}^{1/2}\Delta T_{st}^{-1/2}$ and $\Psi_n \sim \tau_{\mathrm{max}}$, where $\Delta T_{st}$ represents the temperature range of subthermocline isotherms. In the diffusively controlled limit, the classical scalings are recovered: $\delta_{ST} \sim \kappa_b^{1/3}$ and $\Psi_n \sim \kappa_b^{2/3}$, with negligible dependence on wind stress. Because of the constancy of the streamfunction along subthermocline isotherms, the scaling of the cross-equatorial flow, $\Psi_e$, mirrors that of $\Psi_n$ in the adiabatic limit, highlighting the efficiency of adiabatic dynamics in maintaining hemispheric overturning asymmetry. By contrast, in the diffusive regime $\Psi_e \sim \kappa_b^{0.4}$, confirming that enhanced mixing weakens the hemispheric asymmetry of the GOC.
	
	Two analogous limits exist for the scaling of abyssal stratification, $\delta_A$, and the associated AOC, $\Psi_a$. In the adiabatic limit, where eddy and wind-driven effects largely cancel and dominate over diffusive basin dynamics, the scalings are $\delta_A \sim \tau_{\mathrm{max}}^{1}$ and $\Psi_a \sim \tau_{\mathrm{max}}^{-1}$, with no dependence on $\kappa_b$ for $\delta_A$ and $\Psi_a\sim\kappa_b^{1}$. In the diffusively controlled limit, the scalings are $\delta_A \sim \kappa_b^{1/2}$, $\Psi_a \sim \kappa_b^{1/2}$, with negligible dependence on wind stress ($\delta_A \sim \tau_{\mathrm{max}}^0$, $\Psi_a \sim \tau_{\mathrm{max}}^0$). Notice that for both limits, when $\kappa_b \to 0$, $\Psi_a \to 0$, in agreement with the RGGOCM, where the AOC is a purely diffusive cell sustained by eastern boundary diapycnal upwelling.
	
	The scaling behavior of the RGGOCM aligns with theoretical and numerical results from three-dimensional GCMs \citep{ito2008control,wolfe2011adiabatic,nikurashin2011theory,nikurashin2012theory}. However, the RGGOCM offers a transparent derivation, as its geostrophic formulation directly links stratification, zonal asymmetries, and overturning strength. While recently developed column models can reproduce similar scaling laws \citep{nikurashin2012theory,jansen2019toy}, their columnar geometry limits the range of representable forcing scenarios. Furthermore, when the influence of adiabatic channel dynamics on the basin is reduced, the RGGOCM produces a basin overturning streamfunction with two thermocline extrema, resembling the enclosed double-hemisphere configuration (Figs.~\ref{F:DH_MOC}, \ref{F:Channel_kv_scaling} and \ref{F:Channel_tau_scaling}). Such flow structures can not be captured by column models. Under climate change, the loss of overlapping isopycnals connecting the channel to the NH may further diminish the impact of adiabatic channel dynamics on the basin \citep{wolfe2015multiple,vanwesten2025JGR}. Consequently, the RGGOCM provides a more physically consistent framework for representing the basin overturning circulation under extreme forcing.
	
	Although the RGGOCM relies on assumptions that are redundant in fully three-dimensional models, it offers several advantages. Its reduced dimensionality (1) facilitates interpretation of transient and equilibrium behavior and (2) drastically lowers the computational cost of long simulations. This efficiency is particularly valuable for studying the GOC under extreme forcing or investigating multi-stable overturning states, which require extended quasi-equilibrium runs to probe tipping behavior and feedbacks \citep{van2023asymmetry,vanderborght2025feedback}. Moreover, the RGGOCM provides a convenient platform for testing eddy and mixing parameterizations and their impact on high-latitude sinking before implementation in comprehensive GCMs.
	
	To study tipping behavior, a natural extension of the RGGOCM would be a prognostic salinity equation. Combined with relaxation boundary conditions, this could allow the model to exhibit multiple steady states \citep{dijkstra2005stability}. Another extension would be to represent the flow in multiple basins, which can modify classical scaling relations \citep{ferrari2017model,nadeau2020overturning,baker2020meridional} and, under extreme forcing, influence the likelihood of severe mid-depth cell weakening \citep{baker2025continued}.

	\section*{Acknowledgments} 
	E.Y.P.V. and H.A.D. were funded by the European Research Council through the ERC-AdG project TAOC (project 101055096, PI: Dijkstra). The authors gratefully acknowledge Jörn Callies for valuable discussions on the two-plane model, Malte Jansen for assistance in configuring the MITgcm simulations, and Maxim Nikurashin for clarifying the implementation of the boundary-value problem (\ref{psic_edd}).
	
	\section*{Availability Statement}
	The model code and analysis scripts will be made available on Zenodo upon
	publication.

	\section*{Appendix A: Numerical implementation}
	The model equations (\ref{Te}), (\ref{Tw}), and (\ref{Tc}) are discretized on an Arakawa C-grid with a uniform vertical grid and a non-uniform meridional grid, using a vertical spacing of $\Delta z = 80$~m and a meridional spacing of $\Delta \theta = 2^\circ \cos(\theta)$. The cosine factor reduces resolution near the equator while increasing it in convective regions. This coarsening in the equatorial zone substantially improves numerical stability, and the results are only weakly sensitive to grid resolution.
	
	The non-dimensional equations are solved using a time-splitting approach \citep{callies2012simple}: horizontal diffusion is treated with a fully implicit scheme, vertical diffusion with a Crank–Nicolson scheme, and convection with a convective adjustment scheme \citep{rahmstorf1993fast}. Advective terms are integrated using a second-order Runge–Kutta (RK2) explicit scheme. In the channel configuration, the advective flux is written in flux-limited form, for which we apply a Van Leer limiter. The time-step size is set to $\Delta t=0.25$ days.
	
	All simulations, except for the spin-up simulations described in Section~\ref{S:enclosed}\ref{S:DH_Adjust} and Section~\ref{S:Channel}\ref{S:Channel_Adj} are initialized from rest. The initial temperature field is prescribed as
	$T_w(z,\theta) = T_e(z,\theta) = T_s(\theta)\exp(-z/\delta)$ and $T_c(z,\theta) = T_s(\theta)\exp(-z/\delta)$, with $\delta = 40$~m, 
	where $T_s(\theta)$ was given in equation (\ref{Ts}). 
	
	\section*{Appendix B: MIT General Circulation Model}
	The MITgcm is configured in the domains described in Section~\ref{S:enclosed}\ref{S:enclosed_Domain} and Section~\ref{S:Channel}\ref{S:Channel_Domain}, with a horizontal resolution of $2^\circ \times 2^\circ \cos(\theta)$ and 30 unevenly spaced vertical levels. Layer thickness varies from 20 m at the surface to 200 m at depth. As before, the meridional resolution is scaled with $\cos(\theta)$, which we found necessary to suppress unresolved gravity waves that would otherwise fill the model domain (Martin Losch, personal communication). Vertical and horizontal viscosities are set to $1 \times 10^{-3}$ m$^2$ s$^{-1}$ and $1 \times 10^{5}$ m$^2$ s$^{-1}$, respectively. Although the horizontal viscosity is very large, it was required to reduce boundary noise. Importantly, we verified that the overturning circulation remains in dominant geostrophic balance despite the enhanced viscosity.
	
	Eddy-induced transport is determined by solving  a boundary value problem (equation~(\ref{psic_edd})), allowing the model to run with zero horizontal diffusivity. Vertical diffusivity is set to zero throughout the domain, except within $4^\circ$ of a vertical ocean boundary. Following \cite{marotzke1997boundary}, this confines vertical motions to the boundaries and produces zonally flat interior isopycnals. Within this boundary mixing region, vertical diffusivity is set to match the value used in the RGGOCM experiment against which the MITgcm results are compared. Forcing profiles ($T_s$ and $\tau_x$) and other model parameters (Tables~\ref{T:Tab1} and \ref{T:Tab2}) are chosen to be similar to the RGGOCM unless explicitly stated otherwise.
	\bibliographystyle{ametsocV6}
	\bibliography{bibliography.bib}
\end{document}